\documentclass[12pt]{iopart}
\pdfoutput=1
\usepackage{color}
\usepackage{amssymb}
\usepackage[export]{adjustbox}
\usepackage{graphicx,wasysym}

\def\beqra{\begin{eqnarray}}
\def\eeqra{\end{eqnarray}}
\def\beq{\begin{equation}}
\def\eeq{\end{equation}}

\def\vp{\bar{\varphi}}

\def\vp{\varphi}

\def\bx{{\bf{x}}}
\def\bk{{\bf{k}}}
\def\bp{{\bf{p}}}
\def\bq{{\bf{q}}}

\def\bv{{\bf{v}}}

\def\cov{\mathrm{cov}}
\def\bV0{{\bf{V_0}}}
\def\re#1{(\ref{#1})}

\def\agt{~\mbox{\raisebox{-.6ex}{$\stackrel{>}{\sim}$}}~}
\def\alt{~\mbox{\raisebox{-.6ex}{$\stackrel{<}{\sim}$}}~}
\def\bx{{\bf{x}}}
\def\by{{\bf{y}}}
\def\br{{\bf{r}}}

\def\bk{{\bf{k}}}
\def\bp{{\bf{p}}}
\def\bq{{\bf{q}}}

\def\bv{{\bf{v}}}

\def\vp{\varphi}

\def\la{~\mbox{\raisebox{-.6ex}{$\stackrel{<}{\sim}$}}~}
\def\ga{~\mbox{\raisebox{-.6ex}{$\stackrel{>}{\sim}$}}~}

\begin{document}
\begin{flushright}
{\small UMN-TH-3632/17}
\end{flushright}

\title{BAO Extractor: bias and redshift space effects}

\author{Takahiro Nishimichi$^{1,2}$, Eugenio Noda$^{3,4}$, Marco Peloso$^{5}$, Massimo Pietroni$^{3,6}$}
\vskip 0.3 cm
\address{
$^1$Kavli IPMU (WPI), UTIAS, The University of Tokyo, Kashiwa, Chiba 277-8583, Japan\\
$^2$CREST, Japan Science and Technology Agency, 4-1-8 Honcho, Kawaguchi, Saitama, 332-0012, Japan\\
$^3$Dipartimento di Scienze Matematiche, Fisiche ed Informatiche dell'Universit\`a di Parma, Parco Area delle Scienze 7/a, I-43124 Parma, Italy\\
$^4$INFN, Gruppo Collegato di Parma, Parco Area delle Scienze 7/a, I-43124 Parma, Italy\\
$^5$School of Physics and Astronomy, and Minnesota Institute for Astrophysics, University of Minnesota, Minneapolis, 55455, USA\\
$^6$INFN, Sezione di Padova, via Marzolo 8, I-35131, Padova, Italy\\
}

\begin{abstract}
We study a new procedure to measure the sound horizon scale via Baryonic Acoustic Oscillations (BAO). Instead of fitting the measured power spectrum (PS) to a theoretical model containing the cosmological informations and all the nonlinear effects, we define a procedure to project out (or to ``extract'') the oscillating component from a given nonlinear PS. We show that the BAO scale extracted in this way is extremely robust and, moreover, can be reproduced by simple theoretical models at any redshift. By using N-body simulations, we discuss the effect of the nonlinear evolution of the  matter field, of redshift space distortions and of scale-dependent halo bias, showing that all these effects can be reproduced  with sub-percent accuracy. 
We give a one-parameter  theoretical model based on a simple (IR) modification of 1-loop perturbation theory, which reproduces the BAO scale from measurements of halo clustering in redshift space at better than $0.1\%$ level and does not need any external UV input, such as coefficients measured from N-body simulations.

\end{abstract}

\maketitle

\section{Introduction}
BAO as a cosmological standard ruler is firmly established as one of the main observables for present and future surveys \cite{Eis05,Cole:2005sx}. The relevant information contained in the linear power spectrum (PS) of dark matter (DM), namely the comoving sound horizon, can be deformed by a number of nonlinear effects \cite{Eisenstein:2006nj}: the nonlinear evolution of the DM density and velocity fields, redshift space distortions, scale dependent galaxy (halo) bias. All these effects, if not properly taken into account, can hinder the possibility of testing cosmological models against future data, which will bring the PS with an accuracy at the percent level, or below.

Various theoretical approaches have been proposed in the past to attack this problem. The standard approach, which can fully take into account all the nonlinear physics listed above, is N-body simulations. Besides that, in an attempt to design a flexible and numerically less expensive tool, many different approaches have been investigated in the past, mainly building upon standard cosmological perturbation theory (SPT) \cite{PT} and improving it by including the effect of large (IR) bulk matter flows \cite{RPTb,MP07b,RPTBAO,Pietroni08,Bernardeau:2008fa,Taruya:2009ir,Senatore:2014via,Blas:2016sfa,Peloso:2016qdr}, and by giving a more accurate description of mode-coupling with short (UV) scales \cite{Baumann:2010tm,Pietroni:2011iz,Carrasco:2012cv,Manzotti:2014loa,Blas:2015tla,Floerchinger:2016hja,Noda:2017tfh}.

As shown in \cite{Noda:2017tfh} the reliability of the computed PS can be extended to a maximum wavenumber which increases  by a factor up to $\sim 3 (6)$ with respect to 1-loop SPT at redshift $z=0 (1)$, at the expense of introducing one extra parameter (at each redshift) to take into account UV effects. 

The BAO damping in the PS  is due mainly to the random displacement of the galaxies from their original location, by an average of $\sim 6\, {\mathrm{Mpc/h}}$. Reconstruction techniques have been developed \cite{Eisenstein:2006nk,Seo:2008yx,Padmanabhan:2008dd,Noh:2009bb,Tassev:2012hu,White:2015eaa} to undo the effect of these bulk motions and get back a correlation function closer to the linear one, and have been successfully applied to real data (see for instance \cite{Anderson:2013zyy}).
The effectiveness of these techniques, which are based essentially on the Zel'dovich approximation or on variants of it, can be seen as a proof, {\em a posteriori}, that the physics of the BAO peak degradation is well understood and under control \cite{Peloso:2015jua}.

In ref.~\cite{Noda:2017tfh} we proposed a new approach to the extraction of the BAO scale from data, and to the theoretical computation of the related observable. Instead of processing the linear PS to add up all the nonlinear effects and then to fit it to the fully nonlinear PS obtained from observation, we identified a procedure, which we call the ``BAO extractor" (defined in eq.~\re{Rcont}), to project out from the nonlinear PS the oscillating part, containing the BAO information. As we showed in ~\cite{Noda:2017tfh}, the nonlinear evolution of the extracted PS is basically insensitive to the UV effects, while the damping of the oscillations caused by IR flows can be successfully reproduced by a very simple, and theoretically well grounded, analytical procedure. As a result, the time evolution of the extracted BAO's can be followed, at all cosmological epochs down to $z=0$, with subpercent accuracy and with very fast and flexible numerical tools.  

The analysis of ~\cite{Noda:2017tfh} was limited to the DM field in real space. In this paper we extend it to include the effects of redshift space distortions (RSD) and scale-dependent halo bias. In general, the theoretical approaches to these two classes of nonlinear effects are more problematic than for the nonlinear evolution of the DM field. In redshift space, the Fingers of God (FoG) effect, due to virialized motions inside DM halos, cannot be studied in approaches based on the single stream approximation, and therefore one has to resort to a phenomenological modeling. Halo bias, on the other hand, cannot be described entirely in terms of local functions of the DM density field and, again, its description is poorly managed by a SPT approach and by its more straightforward extensions. Therefore, a crucial test of the BAO extractor proposal is the assessment of its robustness against RSD and halo bias effects.

We find that the largest nonlinear effect on the extracted BAO scale is given by the evolution of the DM field, which is above $1~\%$ at $z=0$. However, as we already showed in ~\cite{Noda:2017tfh} and confirm here on more quantitative grounds, this effect can be reproduced by a very simple theoretical model which is basically given by 1-loop SPT plus a resummation of the IR bulk flows (see eq.~\re{Pmodel}). The inclusion of UV effects as discussed in ~\cite{Noda:2017tfh} leads to a further improvement of the BAO scale and of the full PS shape. The other two effects considered in this paper, RSD and bias, are both subpercent level effects to the BAO scale, which are accurately reproduced by our model. Remarkably, all these results are obtained without the need to introduce any nuisance parameter. For comparison, the standard approach to BAO measurements, consists in fitting the full PS with a multi-parametric model (with typically 8 parameters) and fitting functions to separate the oscillating component of the PS.
We also discuss the improvement of our results by including extra parameters to model the nonlinear effects, finding that while it does improve the fit to the full PS, it leads to at most a  moderate gain in the extracted BAO scale.

As a result, our procedure typically leads to a reduced theoretical systematic error  on the measured BAO scale. In the concrete example discussed in this paper the reduction amounts at a factor of order 4-5, but in general it depends on the concrete set of data to which the two procedures are applied.  These results are relevant in light of the forecasted statistical precision of future surveys, such as  DESI and EUCLID,  which will reach subpercent BAO measurements. 

The nonlinear evolution of BAO and the related issue of the robustness of the acoustic scale have been investigated in the past. The effect of IR flows was already identified as the main physical effect to be taken into account (see for instance \cite{Eisenstein:2006nj, Padmanabhan:2008dd,Peloso:2015jua,Anselmi:2017cuq}). In this paper, we explore this issue from the point of view of our BAO extractor procedure, aiming at illustrating its advantages both on the theoretical and on the observational sides.

The plan of the paper is the following. In Section \ref{sec:R[P]} we introduce the procedure to extract the BAO scale from a PS, and the likelihood function to quantify the agreement between the scale obtained in different schemes (or between a given theoretical scheme and the data). In Section \ref{sec:themodel} we introduce a simple model that can reproduce with sub-percent accuracy the BAO scale of halos in redshift space. The model improves over 1-loop SPT by resumming the effect of IR bulk flows, and it does not require to extract any UV parameter from simulations. The robustness of this method against short-scale effects such as the nonlinear evolution of the dark matter field, redshift space distortions, and halo bias is studied in Section \ref{sec:nonilinear}. In Section \ref{sec:conclusions} we present our conclusions. This is followed by four appendices where we, respectively, summarize the TRG method that we use to reproduce the nonlinear evolution of dark matter, perform the resummation of the IR bulk flows in real and redshift space, present some details on the N-body simulations used in this work, and investigate the effect of nondiagonal terms in the PS covariance relevant to our simulations, showing that they are negligible.

\section{BAO extractor: definition}
\label{sec:R[P]}

For any given power spectrum $P\left( k \right)$ we define  \cite{Noda:2017tfh} 
\begin{equation} 
R \left[ P \right] \left( k; \Delta ,\, n \right) \equiv 
\frac{\int_{-{\Delta}}^{\Delta} d x \, x^{2n} \left( 1 - \frac{P \left( k - x\, k_{s} \right)}{P \left( k \right)} \right)}{\int_{-\Delta}^{\Delta}  d x \, x^{2n} \left( 1 - \cos( 2 \pi x )\right)} \;,
\label{Rcont} 
\end{equation} 
where the BAO wavenumber is given by $k_{s}\equiv 2 \pi/r_{s}$, with $r_s$  the comoving sound horizon, computed using, for the assumed cosmological model, eq.~(6) of \cite{Komatsu:2008hk}.
In this expression we integrate around each value of the comoving momentum $k$ in an interval given  by twice the ``range'' parameter $\Delta$ times the BAO wavenumber. 

The operation \re{Rcont}  is similar to
the {\it moving average method}, common in other fields to discern smooth back-ground and rapid oscillations, and has the effect of ``extracting'' the oscillating part of the PS from the smooth one. In \cite{Noda:2017tfh} we showed how the oscillating PS   projected out in this way to a large extent evolves independently under nonlinear effects. 

We also verified that the operation $R\left[ P \right]$, once applied to theoretical power spectra,  is very weakly dependent on the parameter $n$. This parameter might be relevant in analyzing real data, in the case in which the experimental error varies significantly within each interval  $\left[ k -  k_{bao}\,\Delta,  k + k_{bao} \,\Delta \right]$. In the analyses performed in this work we fix $n=0$ and will indicate $R[P](k;\Delta,n=0)$ with $R[P](k;\Delta)$.

LSS power spectra comprise of a smooth broadband (``no-wiggle'') component, plus a smaller (``wiggly'') component due to the BAO oscillations, 
\begin{equation}
P \left( k \right) = P^{nw} \left( k \right) +  P^{w} \left( k \right) \simeq  P^{nw}(k) \left[ 1 + A(k) \sin(k \,r_{bao}) \right] \,, 
\label{PSsplit}
\end{equation} 
where $A \left( k \right) $ is a smooth modulating function which damps the oscillations beyond the Silk scale. While $r_s$ is the BAO scale of the assumed (reference) cosmological model, which is needed to define the extractor (eq.~\re{Rcont}) and the various theoretical formula, as for example, eq.~\re{Pmodel}, $r_{bao}$ is the true scale of the BAO oscillations in data. Different computational techniques reproduce the $ P^{nw} \left( k \right) $ with different accuracy, and for instance a percent accuracy at $k \ga {\rm O }  \left( 0.1\, {\rm h / Mpc } \right)$ requires going beyond standard perturbation theory, and accounting for UV effects through methods such as Coarse Grained Perturbation Theory \cite{Pietroni:2011iz, Manzotti:2014loa} or the Effective Field Theory of LSS \cite{Baumann:2010tm, Carrasco:2012cv}. 

\begin{figure}
\includegraphics[width=0.8\textwidth,right]{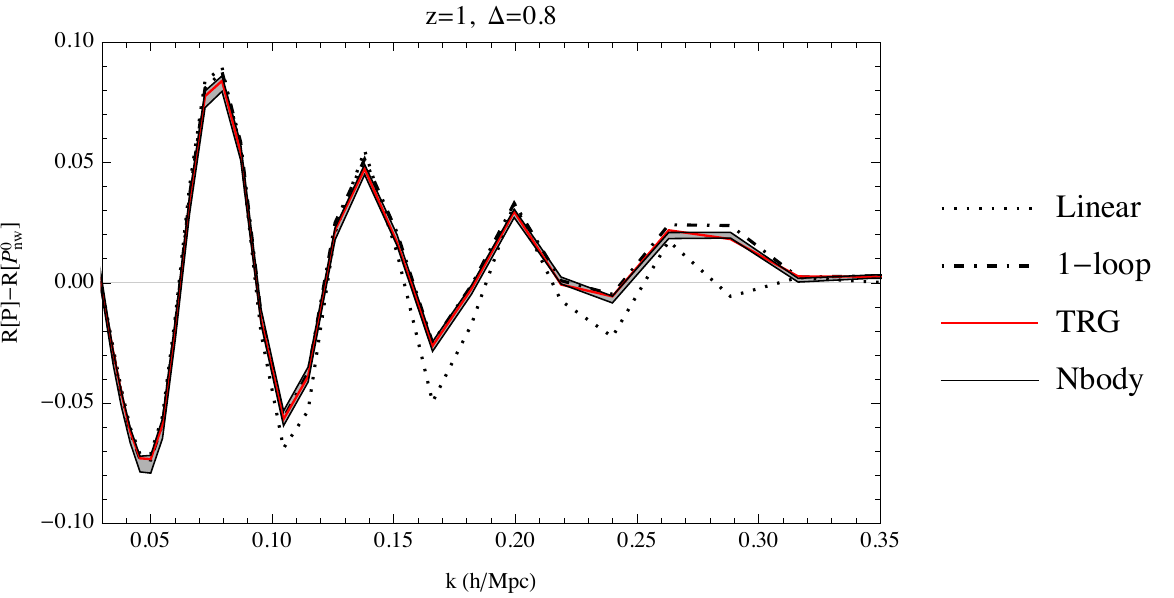}
\includegraphics[width=0.8\textwidth,right]{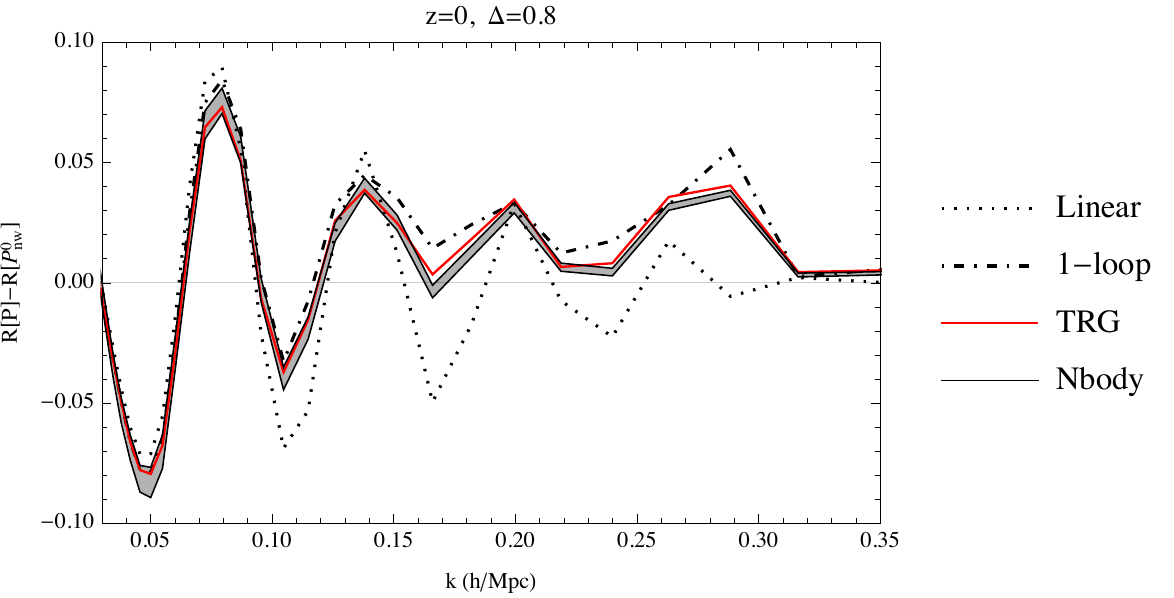}
\caption{The extractor $R[P](k;\Delta)$ applied to the DM PS in real space computed from N-body simulations (grey area), from linear theory (blue, dashed), 1-loop SPT (black, dash-dotted), and TRG (red,  obtained from eq. (\ref{Ptrg})), at redshift $z=1$ (upper panel) and at $z=0$ (lower panel). The grey area for the N-body results corresponds to the diagonal entries of eq.~\re{DRdisc} and using the $1 \sigma$ error on the PS for each bin. For visualisation purposes, the same quantity $R[P^{0,nw}](k)$, where $P^{0,nw}$ is the smooth component of the linear PS, has been subtracted from all the different $R[P](k)$. }
\label{newfig6}
\end{figure}

Inserting \re{PSsplit} in \re{Rcont} for $k_s\,\Delta \ll k$ gives \cite{Noda:2017tfh}
\beq
R[P](k;\Delta)=
I\left(\frac{r_{bao}}{r_s};\Delta\right)\frac{P^w(k)}{P(k)}\; + O\left(1/ \left( k \, r_{s} \right)^2\right)\,,
\label{Rw}
\eeq 
where the scale independent quantity $I(\beta;\Delta,n)$ is given by
\beq
I\left(\beta;\Delta\right) \equiv 
\frac{\int_{-\Delta}^{\Delta}  d x \,  \left( 1 - \cos\left( 2\pi\, \beta\, x \right)\right)}{\int_{-\Delta}^{\Delta}  d x \, \left( 1 - \cos( 2 \pi x )\right)}\,,
\eeq
and $I\left(1;\Delta\right) =1$.
 Notice that the quantity  $k_{s}$ (and therefore $r_{s}$) enters only to set the appropriate units of the interval over which the integral in \re{Rcont} has to be taken. It does not affect the scale of the oscillations  of $R[P]$ such as those in  Fig.~\ref{newfig6}, which is given by the $r_{bao}$ contained in $P^w(k)$ (see eq.~\re{PSsplit}). Choosing $r_s\neq  r_{bao}$  only affects -- typically very mildly-- the amplitude of the oscillations (see Fig.~4 of  \cite{Noda:2017tfh}). 
 
As shown in \cite{Noda:2017tfh} and in the present work, the scale of the oscillation that emerges from $R \left[ P \right]$, is very weakly sensitive to the details of the smooth $ P^{nw} \left( k \right) $ component. This allows to extract the BAO scale from the data, and to compare it with theory, in a way which is very insensitive to the UV physics. Details on how this comparison is performed are given below, see eq. (\ref{chi2}). 

The expression (\ref{Rcont}) is suitable for an input continuous power spectrum as given by theory. Assume we have instead binned data with 
 momentum / PS value / error on the PS given by $\left\{ k_n ,\, P_n ,\, \Delta P_n \right\}$. From the data we can construct the estimator  \cite{Noda:2017tfh} 
\beqra
&&\!\!\!\!\!\!\!\!\!\!  \hat R[P](k_m;\Delta) \equiv \frac{\sum_{l=-L_m \left( \Delta \right)}^{L_m \left( \Delta  \right) } \, \left(1-\frac{P_{m+l}}{P_m}\right)}{\sum_{l=- L_m \left( \Delta \right)}^{L_m \left( \Delta \right)} \left[ 1-\cos\left(r_s \left(k_{m+l}-k_m \right)\right)\right]}\,,
\label{Rdisc}
\eeqra
where the value of the maximum $|l|$ in the sum, $L_m \left( \Delta \right)$, is chosen such that
\beq 
\left| k_{m+l}-k_m \right| \le k_{s} \, \Delta  \;\;\;\; {\mathrm{for}}\;\; \vert l \vert \le L_m \left( \Delta \right)\,.
\eeq

Under the assumption of uncorrelated errors for the PS across different bins, the covariance matrix of the estimator is given by 

\beqra
\!\!\!\!\!\!\!\!\!\! \!\!\!\!\!\!\!\!\!\!\!\!\!\!\!\!\!\!\!\! &&C_{n,m}(\Delta)\equiv \langle  \hat R[P](k_n;\Delta) \hat R[P](k_{m};\Delta)\rangle - \langle\hat R[P](k_n;\Delta)\rangle \langle \hat R[P](k_{m};\Delta)\rangle \nonumber\\
&&=D_m^{-1}D_n^{-1} \sum_{l=-L_m(\Delta)}^{L_m(\Delta)}\sum_{l'=-L_n(\Delta)}^{L_n(\Delta)} \frac{P_{m+l}P_{n+l'}}{P_{m}P_{n}}\Bigg(\frac{\cov^P_{m,n}}{P_{m}P_{n}}+\frac{\cov^P_{m+l,n+l'}}{P_{m+l}P_{n+l'}}\nonumber\\
&&\qquad\qquad \qquad-\frac{\cov^P_{m+l,n}}{P_{m+l} P_{n}}-\frac{\cov^P_{m,n+l'}}{P_{m}P_{n+l'}}\Bigg)\,,
\label{DRdisc}
\eeqra
where
\beq
D_n\equiv \sum_{i=-L_n \left( \Delta \right)}^{L_n \left( \Delta \right)} \left(1-\cos\left(r_s \left(k_{n+i}-k_n \right)\right)\right)\,,
\eeq
and $\cov^P_{mn}$
is the covariance matrix for the PS.

In the rest of this paper, when considering  different PS's $P_a$, we will plot the subtracted quantities $\hat R[P_a]-\hat R[P^{0,nw}]$, where $\hat R[P^{0,nw}]$ obtained by applying the procedure \re{Rdisc} to the smooth linear PS $P^{0,nw}$, defined as explained in \ref{appendicetrg}. This subtraction largely removes the smooth $O(1/(k r_s^2))$ contribution to eq.~\re{Rw}. However, in our analyses, we always consider the unsubtracted $\hat R[P_a]$'s. Moreover, in the following, we will indicate the estimator $\hat R[P_a](k_m;\Delta)$ simply as $R[P_a](k;\Delta)$.

We plot in Fig.~\ref{newfig6} the result of the extraction procedure applied to different approximations of the DM PS in the BAO range of scales, namely, linear theory, 1-loop SPT, improvements to 1-loop SPT discussed in the next two sections, and N-body simulations. This plot will be further discussed in subs.~\ref{nonlincorr}.

Besides damping the oscillations with respect to linear theory, nonlinearities also modify the BAO scale, an effect which can crucially hinder the possibility of using BAO's as a cosmic ruler \cite{Sherwin:2012nh} (although, as we mentioned before, this effect can be reduced by reconstruction techniques). In order to quantify this effect we define a likelihood function as
\beq
\chi^2_a(\alpha)=\sum_{n,m=n_{min}}^{n_{max}}\delta R[P_a](k_n;\alpha,\Delta)C^{-1}_{n,m}(\Delta)\delta R[P_a](k_m;\alpha,\Delta)\,, 
\label{chi2} 
\eeq
where
\beq
\delta R[P_a](k_n;\alpha,\Delta)\equiv R[P_a](k_n/\alpha;\Delta) -R[P_{data}](k_n;\Delta)\,,
\eeq
$P_a$ the model PS as obtained in a given approximation, while $P_{data}$ is the measured one, and $C^{-1}_{n,m}(\Delta)$ is the inverse of the matrix given in eq.~\re{DRdisc}, computed using the experimental errors on the data PS. The sum is taken over the momenta $k_n$ in which there are BAO oscillations (see Fig.~\ref{newfig6}),  typically in the range
\beq
0.025 \;\;{\mathrm{h\,Mpc^{-1}}} \alt k_n\alt 0.3 \;\;{\mathrm{h\,Mpc^{-1}}}\,,
\eeq
where the upper end is limited by the goodness of the $\chi^2_a(\alpha)$ value, and therefore varies for the different models/approximations. For those models providing a good fit up to higher momenta, the $\chi^2_a(\alpha)$ curve is narrower and therefore gives tighter constraints on the extracted BAO scale.

In the following, we use the diagonal entries of the covariance matrix for the extractor, eq.~\re{DRdisc}, to indicate the $1\sigma$ errors on the extracted $R[P]$ functions in plots such as those in Fig.~\re{newfig6}, while we use the full matrix to evaluate the various $\chi_a^2(\alpha)$. As for the covariance matrix for the PS from N-body simulations, $\cov^P_{m,n}$, we will estimate its diagonal terms from the scattering of $|\delta_\bk|^2$ with $\bk$'s inside each $k$ bin, while we will neglect the nondiagonal terms. In \ref{COVNB} we will investigate the impact of the non-diagonal terms on our analyses, showing that they are negligible.

\section{A simple model for the extraction of the BAO scale}
\label{sec:themodel}

In this section we present a simple procedure to extract the BAO scale from a given $P_{data}$. We will use, as data, the halo PS in redshift space (see \ref{Simul} for technical details on our simulations and halo catalogs). The impact of the different types of nonlinear effects on $R[P]$, and the performance of different approximation methods in dealing with them will be discussed in the next sections.

Our model PS is given by
\beq
P_{model}(k,\mu;A)=e^{-A k^2} P_{res}(k,\mu)\,,
\label{Pmodel}
\eeq
with
\beqra 
&& P_{res}(k,\mu) = P^{nw,rs,0}(k,\mu) + \Delta P^{nw,rs,1-loop}(k,\mu)\nonumber\\
&&\qquad\qquad\quad+  P^{w,rs,0}(k,\mu) \,e^{-k^2\Xi^{rs}(\mu;r_{s})}\,,
\label{PtrgRS}
\eeqra
where $P^{nw,rs,0}(k,\mu)$ and $P^{w,rs,0}(k,\mu)$ are the smooth and ``wiggly" components of the linear PS for DM in the Kaiser approximation, namely, 
\beq
P^{a,rs,0}(k,\mu)=(1+\mu^2 f)^2P^{a,0}(k)\,,
\eeq
with $a=w,nw$.
The 1-loop correction in the improved Kaiser approximation is given by
\beqra
&&\Delta P^{nw,rs,1-loop}(k,\mu)=\Delta P_{\delta\delta}^{nw,1-loop}(k)+2 \mu^2 f \Delta P_{\delta\theta}^{nw,1-loop}(k)\nonumber\\
&&\qquad\qquad\qquad\qquad + \mu^4 f^2 \Delta P_{\theta\theta}^{nw,1-loop}(k)\,,
\eeqra
with $\Delta P_{\delta\delta}^{nw,1-loop}(k)$, etc, the different components of the real space 1-loop PS \cite{PT} for the density contrast $\delta$ or the velocity divergence $\theta$ computed from the linear smooth one.

The resummation of the effect of IR random velocity flows at all orders in SPT  is implemented by exponentiating 
\beqra
&&\Xi^{rs}(\mu;r_{s})= \left(1+f\mu^2(2+f)\right)\Xi(r_{s})\nonumber\\
&&\qquad\qquad\qquad\qquad +f\mu^2(\mu^2-1)\frac{1}{2\pi^2}\int dq\,P^{nw,0}(q) j_2(qr_{s})\,,
\label{XiRS}
\eeqra
with
\beqra
&&
\Xi(r_{s}) \equiv \frac{1}{6 \pi^2}\int  dq\,P^{nw,0}(q;z)\,( 1-j_0(q\,r_{s}) + 2 j_2(q \,r_{s}))\,,
\eeqra
see \ref{appendicetrg} and \cite{Noda:2017tfh}, where the details of the resummation procedure are discussed.

Multipoles can be computed from $P_{res}(k,\mu)$ as usual, by taking the integrals
\beq
P_{model,l}(k;A)\equiv \frac{2l+1}{2}\int_{-1}^1d\mu\; P_{model}(k,\mu;A)\,\mathcal{P}_l(\mu)\,,
\eeq
where $\mathcal{P}_l(\mu)$ is the Legendre polynomial of order $l$. In what follows we will only consider the monopole ($l=0$).

\begin{figure}
\includegraphics[width=0.8\textwidth,right]{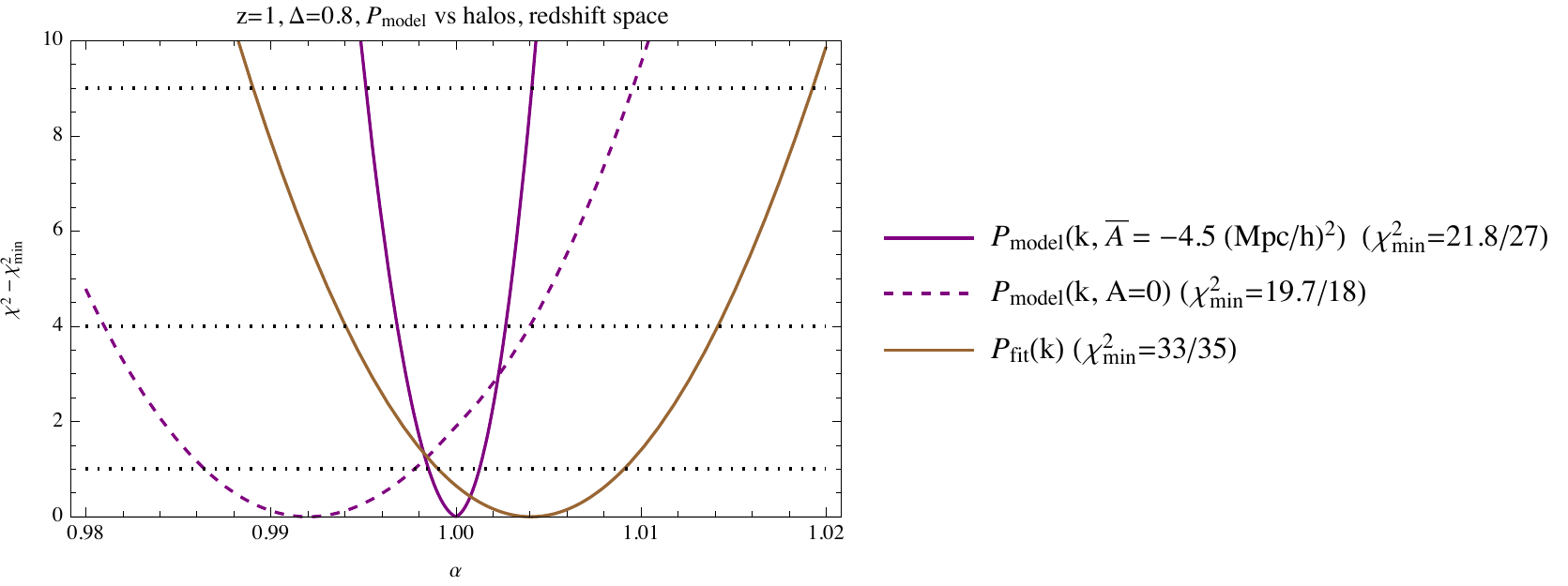}
\includegraphics[width=0.8\textwidth,right]{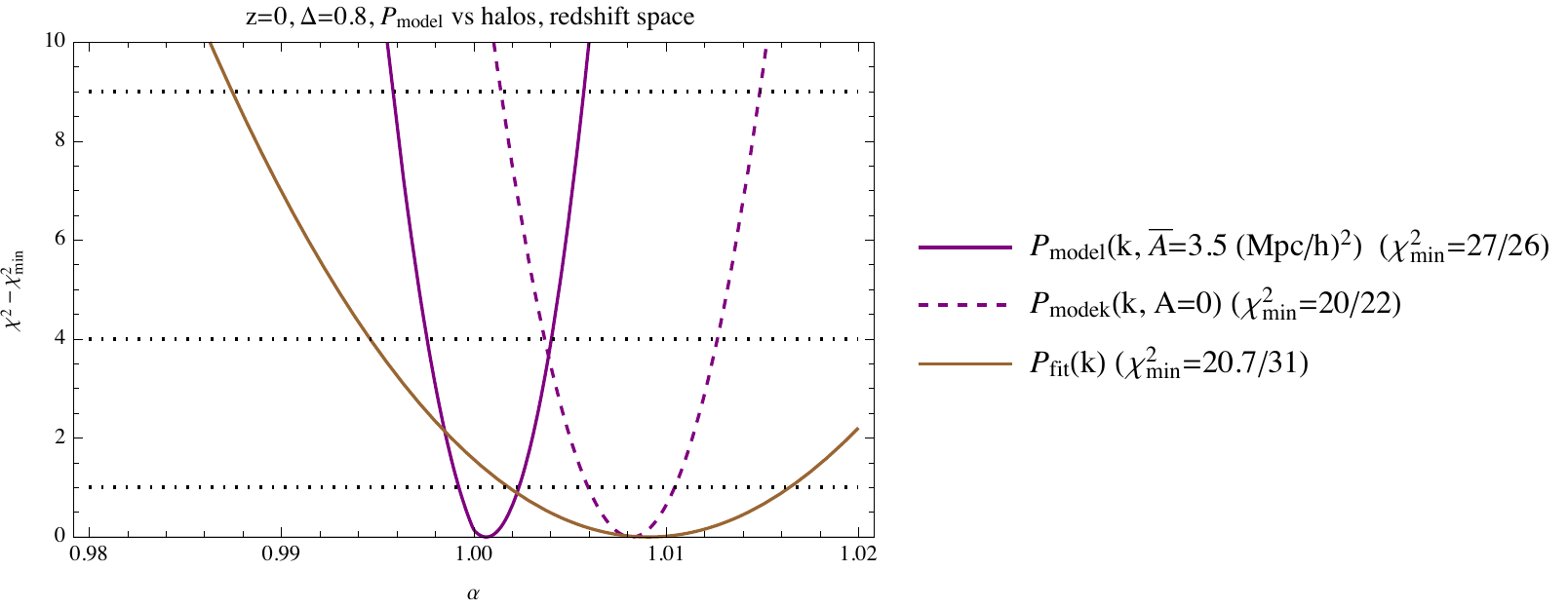}
\caption{The function $\chi^2$ defined in \re{chi2} as a function of the shift parameter $\alpha$ with respect to halo data  using $P_{model,0}(k;A)$ as a model. Each curve has been shifted vertically so to be vanishing at its minimum. The effect of removing the free parameter $A$ is shown by the comparison of the purple solid and dashed lines.}
\label{BaoExtr}
\end{figure}

The only free parameter in eq.~\re{PtrgRS} is the constant $A$ at the exponent, which, as we will see, can be treated as a nuisance parameter to marginalize over the scale dependence of halo bias and on redshift space effects not captured by the (improved) Kaiser approximation encoded in $P_{res}(k,\mu)$, such as FoG. In \cite{Desjacques:2008jj,Desjacques:2009kt,Baldauf:2014fza}, a scale-dependent halo bias given by a constant plus a quadratic term in $k$ was predicted by the peaks model, which can be seen as a truncated Taylor expansion of the model used here. Notice that since the definition of $R[P]$, eq.~\re{Rdisc}, is insensitive to the PS normalization, a constant bias parameter would drop off the analysis, so we do not need to introduce it. Moreover, in order to fully capture the unaccounted redshift effects such as FoG, a $\mu$-dependent function should be introduced  (FOG should scale as something like $\exp(-f^2 \sigma_v^2k^2\mu^2))$, see for instance \cite{Scoccimarro:2004tg}. However, in this  paper we only consider the monopole, therefore we decided to minimise the number of extra functions and incorporate both bias and RSD in the single parameter $A$.

To fix $A$ we use the full monopole power spectra $P_{model,0}$ and $P_{halo,0}$. We take the ratios between the two, and look for the value $\bar A$ which minimizes its scale dependence at low $k$'s. Then we compute \re{chi2} as a function of the parameter $\alpha$ with $P_a=P_{model,0}(k,\bar A)$ and $P_{data}=P_{halo,0}(k)$. 
The result is shown in Fig.~\ref{BaoExtr}, where we show the function $\chi^2(\alpha)-\chi^2(\alpha_{min})$ obtained by this procedure. We also show the effect of removing the parameter $A$ by setting it to zero in \re{Pmodel}. As we see, the simple model considered here captures the correct BAO scale at better than the $ 0.1\%$ level both at $z=0$ and at $z=1$, moreover the $1 \sigma$ confidence level corresponds to $0.16\,\%$ ($0.14\,\%$) at $z=0$ ($z=1$). Setting the exponential prefactor to unity ($A=0$), still reproduces the BAO scale at subpercent level, but with a reduced precision, especially at $z=1$, where the effect of halo bias is larger, see Sect.~\ref{sec:halobias}.

For comparison, we also plot (brown solid line) the $\chi^2$ obtained by a different procedure, which resembles more closely the standard one employed, for instance, in \cite{Anderson:2013zyy}. In this case, the model PS is fit directly to the full $P_{data}$ PS, instead of considering the $R[P]$ operation, therefore, more parameters are needed to model the broadband feature. Following \cite{Anderson:2013zyy}, we consider a model PS given by the $8-$parameter function: 
\begin{equation}
P_{\rm fit} \left( k;\alpha \right) = P^{\rm smooth} \left( k \right) \left\{ 1 + \left[ O^{\rm linear} \left( \frac{k}{\alpha} \right) - 1 \right] 
e^{- \frac{k^2 \Sigma_{\rm nl}^2}{2}} \right\} \;\;, 
\label{Pfit}
\end{equation} 
where the oscillatory component is obtained from the ratio between the total linear and smooth linear power spectrum\footnote{In \cite{Anderson:2013zyy} the smooth PS is derived using the fitting formula of \cite{Eisenstein:1997ik} instead of the procedure described in \ref{appendicetrg}. This difference does not change things appreciably. }
\begin{equation}
 O^{\rm linear} \left( k \right) \equiv \frac{P^0 \left( k \right)}{P^{0,{ nw}} \left( k \right)} \,, 
\end{equation} 
while the smooth component is given by
\begin{equation} 
P_{\rm smooth} \left( k \right) \equiv B_P^2 P^{0,{ nw}} \left( k \right) + A_1 \, k + A_2 + \frac{A_3}{k} + \frac{A_4}{k^2} + \frac{A_5}{k^3} \,,
\end{equation} 
where the parameters $A_{1,\cdots,5}$ and $B_P$ marginalise over broad-band effects including redshift-space distortions and scale-dependent bias. Notice that, unlike our $\Xi$ function of eq.~\re{XiRS}, now also the  exponential damping containing $\Sigma_{\rm nl}$ in \re{Pfit} is treated as a nuisance parameter. 
To quantify the precision with which this procedure can reproduce the BAO scale, we fit $P_{halo}$ with the expression
(\ref{Pfit}), fixing $\alpha=1$ and finding the best fit values for the remaining $7$ parameters $\left\{ \Sigma_{\rm nl} ,\, B_P ,\, A_1 ,\, A_2 ,\, A_3 ,\, A_4 ,\, A_5 \right\}$. With these values fixed \footnote{We checked that the procedure is converging, in the sense that by fixing the initial $\alpha$ to a slightly different ($\pm 1\%$) value and extracting the corresponding parameters gives very similar $\chi^2$ curves.   }, we then compute the likelihood  
\beq
\chi^2_{fit}(\alpha)=\sum_n \frac{\left( P_{\rm fit} (k_n,\alpha) -P_{\rm halo} (k_n)\right)^2}{ \left(\Delta  P_{\rm halo} (k_n)\right)^2}\,, 
\label{chifit} 
\eeq
as a function of $\alpha$. 

As seen from the figure, both methods are able to return the BAO scale at the subpercent level, as a best fit and are in mutual agreement within $1\,\sigma$. The advantage of the $R[P]$ method appears when looking at the width of the likelihood intervals, which, for these ``data'', is reduced with respect to the method based on \re{Pfit}:  it gives a $1\sigma$ error on $\alpha$ of  $0.16\,\%$ ($0.14\,\%$) at $z=0$ ($z=1$) against $1.4\,\%$ ($1\,\%$) by the ``standard'' procedure.

\section{Nonlinear effects on $R[P]$}
\label{sec:nonilinear}

In this section we discuss the sensitivity  of the function $R \left[ P \right] \left( k; { \Delta} \right)$ to various nonlinear effects.

\subsection{Nonlinear evolution of the DM field}
\label{nonlincorr}
Assuming that N-body simulations fully account for DM nonlinearities on the scales of interest for this paper, we discuss how different approximations affect the $R[P]$ operation and the extraction of the BAO scale from it.  Besides linear theory and 1-loop SPT, we will consider the TRG result of ~\cite{Noda:2017tfh}, which can be cast, in real space, in the form
\beq 
P^{TRG}(k) = P_{model}(k;\mu=0,A=0)+ \Delta P^{nw,TRG}(k)\,,
\label{Ptrg}
\eeq
where $P_{model}$ has been defined in \re{Pmodel}.  $\Delta P^{nw,TRG}(k)\equiv D(z)^2 \Delta P^{nw}_{11}(k;\eta)$ is the UV correction  ($\eta\equiv \log D(z)$, $(D(0)=1)$), where $\Delta P^{nw}_{11}(k;\eta)$ solves the TRG system discussed in \cite{Noda:2017tfh},
which encodes the difference between the correct UV behavior, extracted from simulations, and the one of 1-loop SPT  added (see \ref{appendicetrg} for a discussion, and ref. ~\cite{Noda:2017tfh} for full details).

\begin{figure}
\includegraphics[width=0.8\textwidth,right]{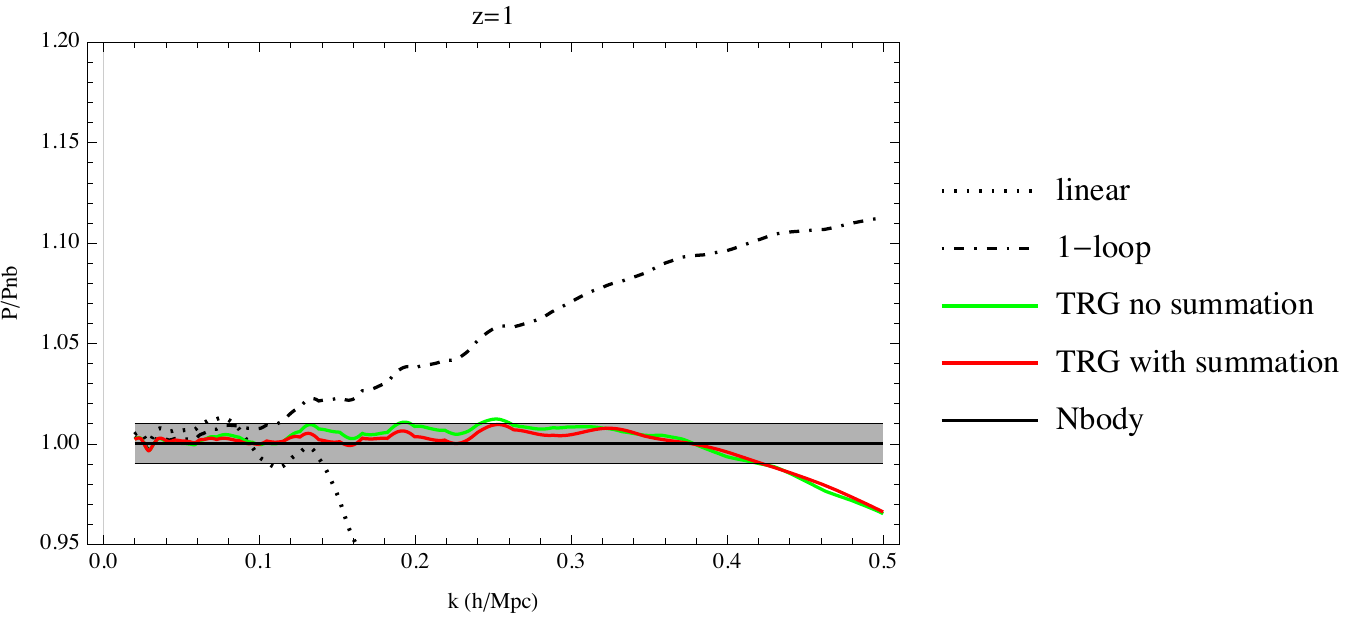}
\includegraphics[width=0.8\textwidth,right]{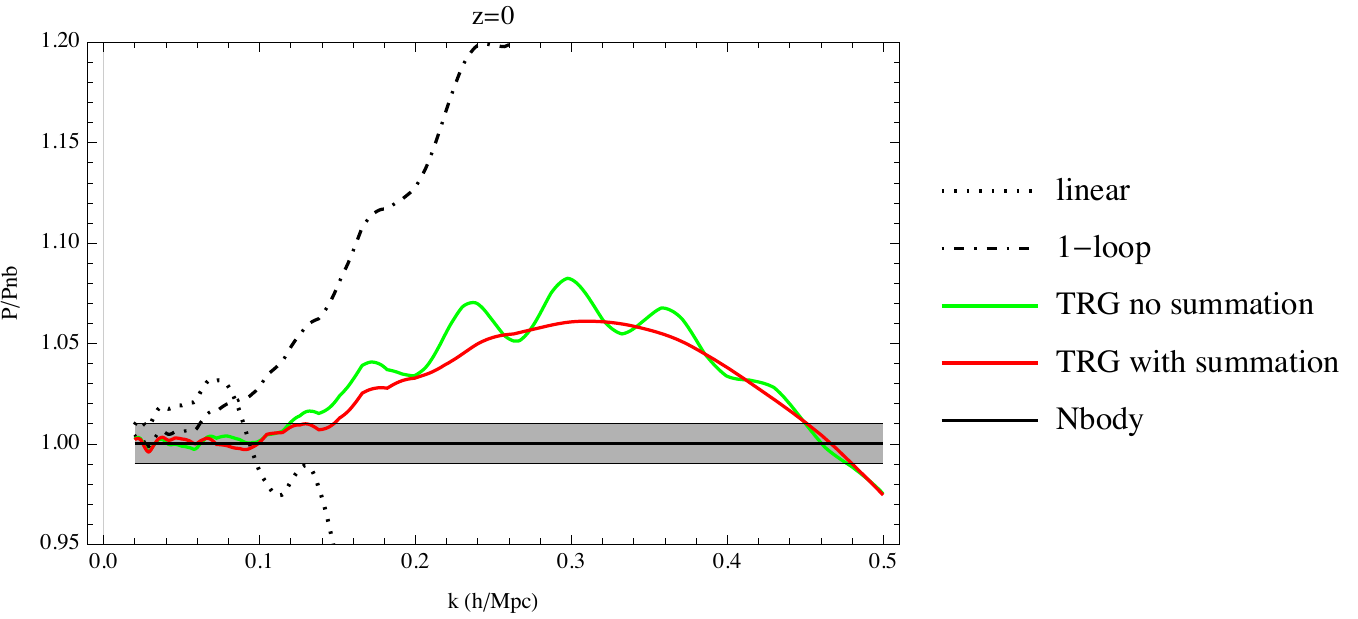}
\caption{Ratios of the nonlinear PS obtained with the different approximations described in the text after eq.~\re{Ptrg} to the N-body ones.}
\label{Pratio}
\end{figure}

To see how well the nonlinear broadband shape of the PS is reproduced, in Fig.~\ref{Pratio} we plot the ratios of the different approximations to N-body simulations. 

In comparison with 1-loop SPT, the UV effects encoded in the TRG corrections increase the $k$-range in which the results agree to better than $1\,\%$ with simulations from $k\sim 0.06\;(0.1)\;\mathrm{h^{-1}\;Mpc}$ to $k\sim 0.14\;(0.42)\;\mathrm{h^{-1}\;Mpc}$ at $z=0$ ($z=1$). On the other hand, the effect of the resummation of the random motions is seen by comparing the red and the green lines, which are  obtained by setting $\Xi^{rs}(0;r_s)=0$ in \re{Ptrg}, and by replacing $ \Delta P^{nw,1-loop}(k)$ with the full 1-loop correction, $ \Delta P^{1-loop}(k)$.

To focus the discussion on the BAO's, we plot in Fig.~\ref{newfig6}, the result of the extraction procedure applied to the different approximations discussed above. For all the approaches we have used the same binning in $k$, and the grey band corresponds to the estimated $1 \sigma$ statistical error on the N-body PS, evaluated using the diagonal entries of eq.~\re{DRdisc}. We see that the TRG approach reproduces the BAO damping quite well over the full range of wavenumbers. 

\begin{figure}
\includegraphics[width=0.8\textwidth,right]{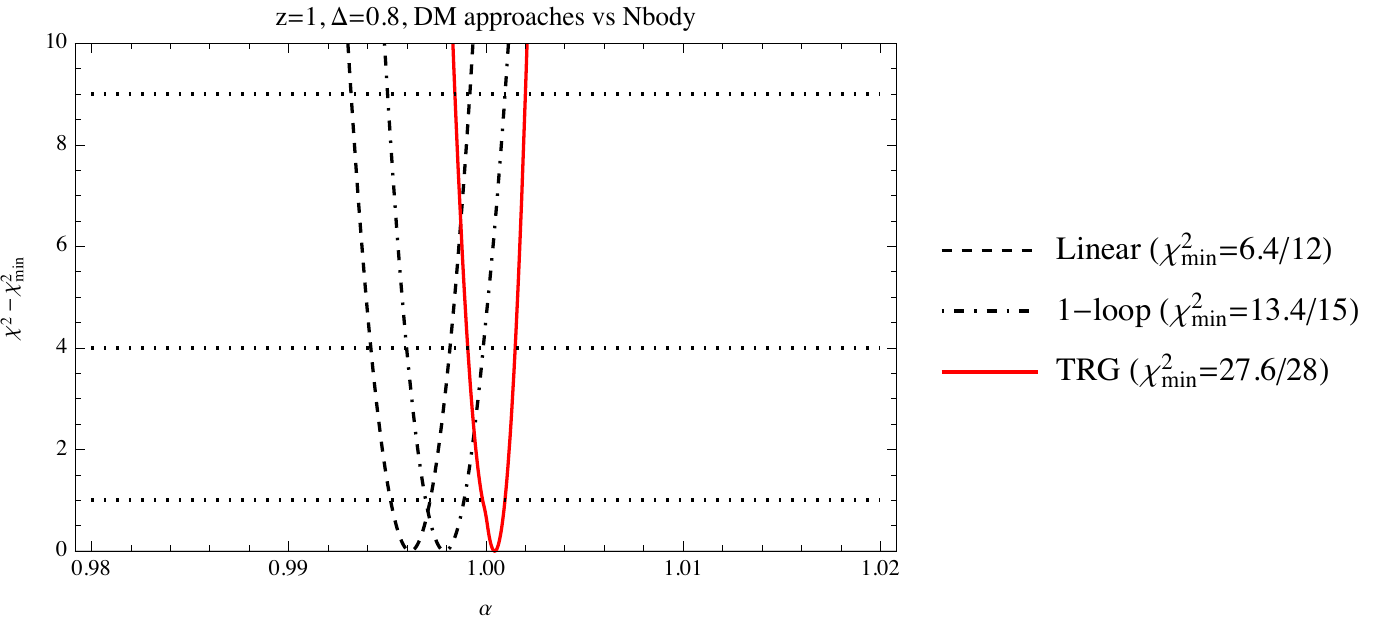}
\includegraphics[width=0.8\textwidth,right]{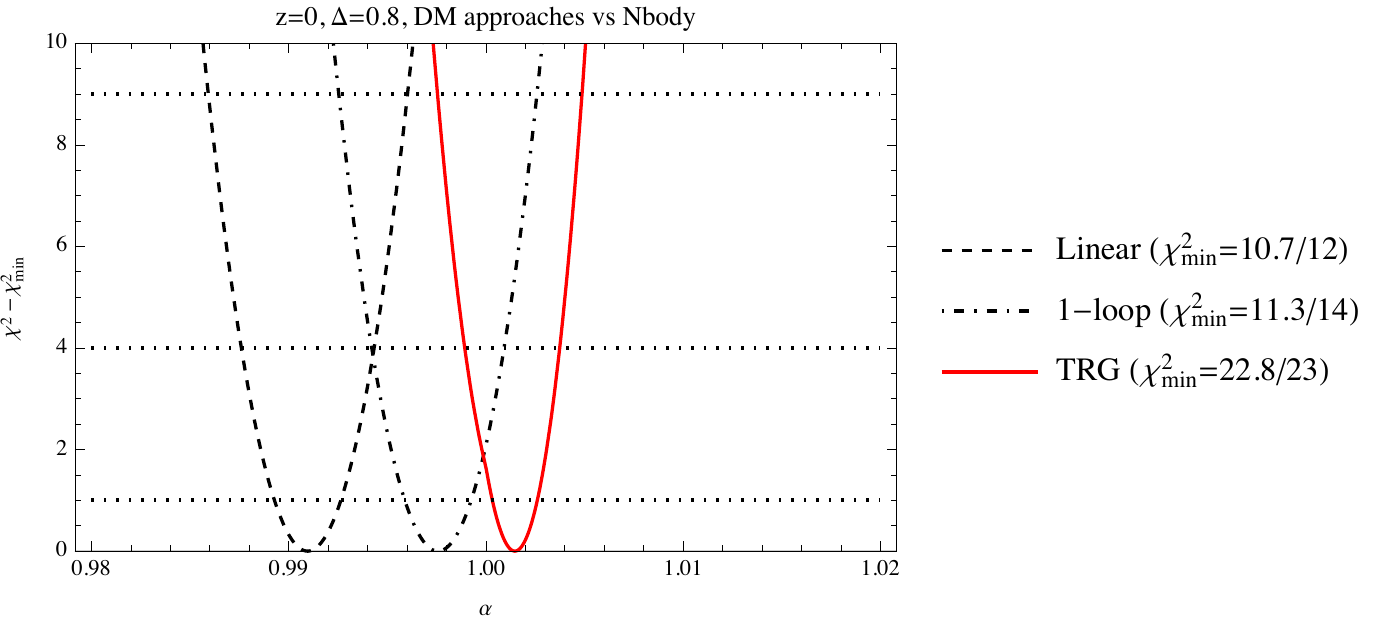}
\caption{The function $\chi^2$ defined in \re{chi2} as a function of the shift parameter $\alpha$ with respect to N-body data  for different approximation to the DM PS. Each curve has been shifted vertically so to be vanishing at its minimum. }
\label{chivsNB}
\end{figure}

To quantify the effect of the nonlinear DM evolution on the BAO scale we define a likelihood function, in analogy to \re{chi2}, in which the role of $P_{data}$ is played by the N-body DM PS, and the covariance matrix is obtained from the statistical errors on the N-body PS , while $P_a$ is obtained in  linear theory, 1-loop SPT, and in the TRG approach.

The results are plot in Fig.~\ref{chivsNB}. The nonlinearities encoded in the N-body simulations shift the BAO scale towards smaller scales by  $\sim 0.8\% \,(0.4\,\%)$ at $z=0$ ($z=1)$ with respect to the linear theory PS \footnote{To get an analytic insight on this effect see \cite{Sherwin:2012nh}. Reconstruction techniques are also able to remove the BAO shift, see for instance \cite{Padmanabhan:2008dd}. }. The 1-loop SPT approximation, on the other hand, gives a much better  fit both at $z=0$ and at $z=1$, mainly thanks to the fact that it reproduces the lower peaks well. It is a remarkable result, as the 1-loop approximation is generally considered a poor tool for the description of the nonlinear PS in the BAO range of scales. This is of course true, but is mainly related to the broadband part of the PS. As far as  the oscillating component extracted by the $R \left[ P \right]$ procedure is concerned, the 1-loop result performs very well. This is fully in line with the previous study by Ref.~\cite{Nishimichi:2007xt}, where it was shown that the BAO scale is robust against the 1-loop SPT correction when the smooth broadband is removed appropriately. We will see, however, that it is not the case in redshift space.

On the other hand,  the TRG result (red lines) reproduces the nonlinear shift at better than $0.2\,\%$ level both at $z=0$ and than $0.1\,\%$ at $z=1$. Given that the computational cost for this approach is the same needed for the 1-loop PS, the gain represented by it at low redshifts is clear.

\begin{figure}
\includegraphics[width=0.8\textwidth,right]{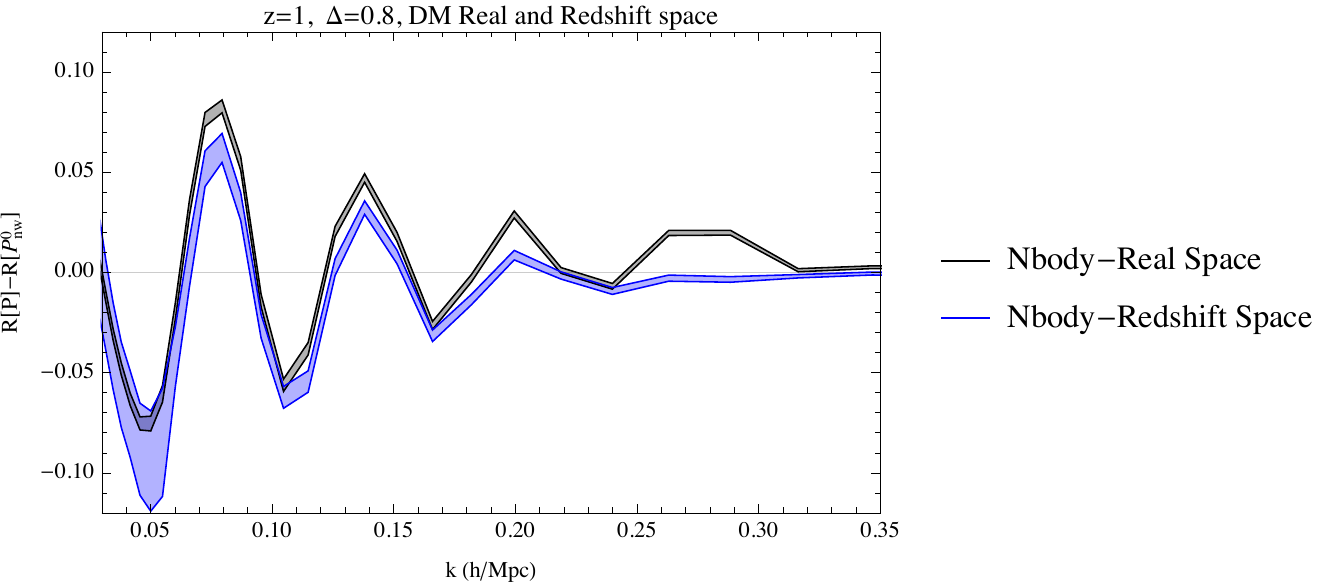}
\includegraphics[width=0.8\textwidth,right]{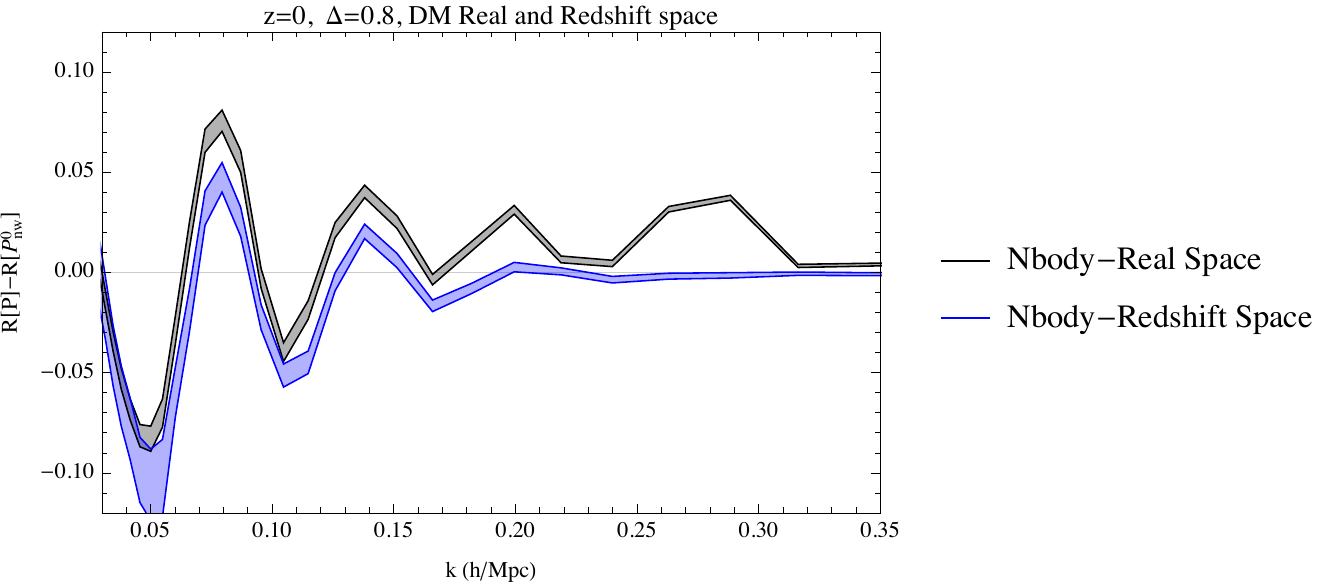}
\caption{Comparison between the $R[P]$'s for the Nbody PS in real and in redshift space (monopole) at $z=1$ and at $z=0$.}
\label{newfig6RS}
\end{figure}

\begin{figure}
\includegraphics[width=0.65\textwidth,center]{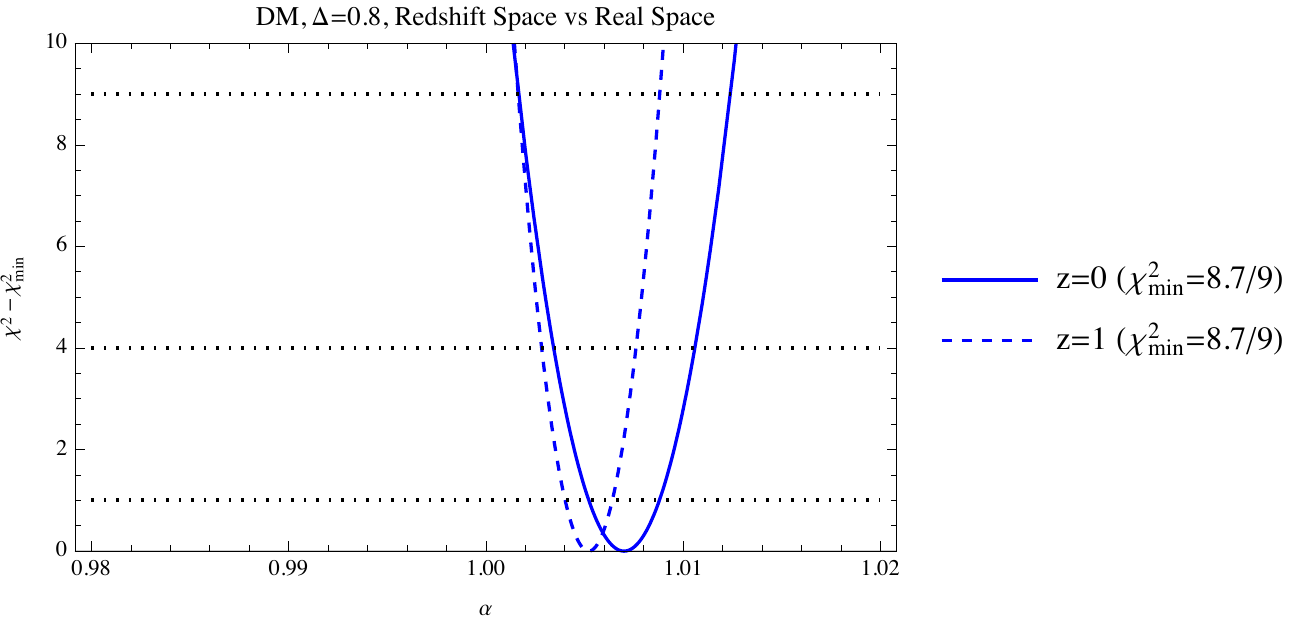}
\caption{$\chi^2$ of the shift $\alpha$ between the real and redshift space PS for DM  from N-body simulations.}
\label{chivsNB2}
\end{figure}

\subsection{Redshift space distortions}

The effect of redshift space distortions on the extracted BAO is given in Fig.~\ref{newfig6RS}, where we plot the results for the DM field as obtained in N-body simulations both in real and in redshift space (in the latter case we consider the monopole PS). The effect on the extracted $\alpha$ parameter is given in Fig.~\ref{chivsNB2}, where we have used again eq.~\re{chi2}, with $P_a$ being the monopole PS in redshift space and $P_{data}$ the N-body PS in real space. In other words, we try to fit the real space $R[P]$ from the redshift space one. As a result, the  extracted BAO scale is rescaled by $\sim 0.5\,\%$, at $z=1$ and by  $\sim 0.7\%$ at $z=0$, but the fit rapidly gets poorer by including higher wavenumbers.  The question is whether one can improve this extraction by applying the TRG approach to redshift space.

The  extension of eq.~\re{Ptrg} to redshift space is (see \ref{appendiceIR})
\beq
P^{TRG,rs}(k,\mu) = P_{model}(k;\mu,A=0)+ \Delta P^{nw,rs,TRG}(k,\mu)\,,
\label{PTRGRSnA}
\eeq
where
\beqra
&&\!\!\!\!\!\! \Delta P^{nw,rs,TRG}(k,\mu)=D(z)^2\Big[ \Delta P_{11}^{nw,TRG}(k,\mu;\eta)+2 \mu^2 f \Delta P_{12}^{nw,TRG}(k,\mu;\eta)\nonumber \\
&& \qquad\qquad\qquad\quad+ \mu^4 f^2 \Delta P_{22}^{nw,TRG}(k,\mu;\eta)\Big]\,,
\label{DPTRGRS}
\eeqra
where $\Delta P_{ab}^{nw,TRG}$ ($a,b=1,2$) are, again, computed using the TRG equations of \cite{Noda:2017tfh}.

\begin{figure}
\includegraphics[width=0.8\textwidth,right]{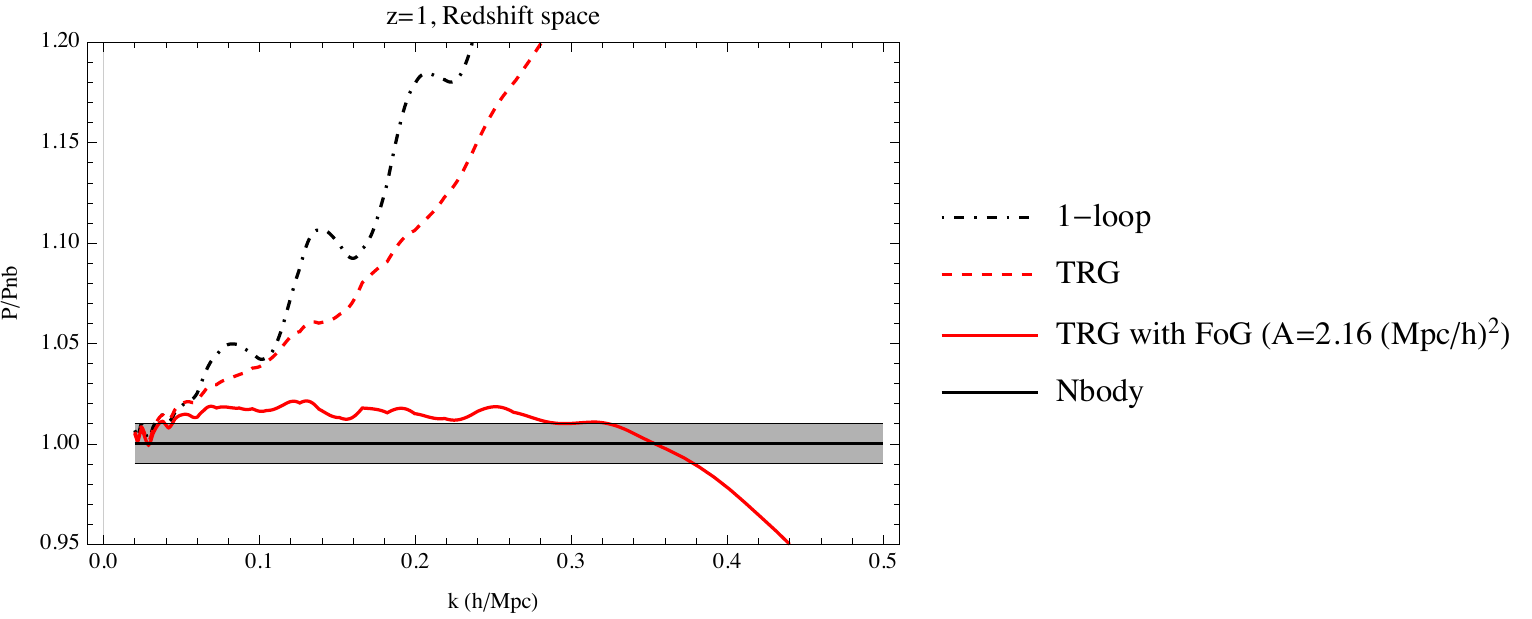}
\includegraphics[width=0.8\textwidth,right]{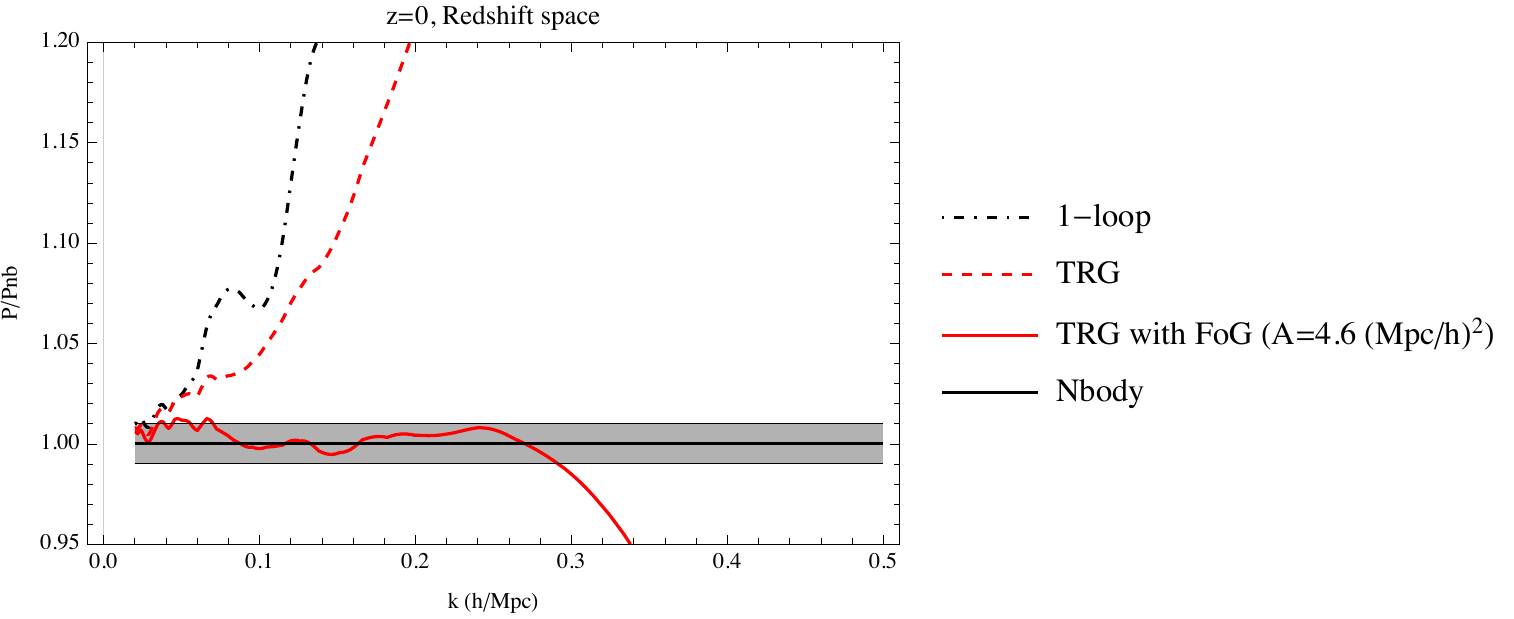}
\caption{ Ratios of the nonlinear (monopole) PS in redshift space obtained with the different approximations described in the text at $z=1$ and at $z=0$.}
\label{PratioRS}
\end{figure}

The performance of eq.~\re{PTRGRSnA} in reproducing the broadband shape of the nonlinear PS in redshift space is much worse than for its real space counterpart, as we can see comparing the red dashed lines of Fig.~\ref{PratioRS} to the red solid ones in Fig.~\ref{Pratio}. It still improves somehow with respect to the 1-loop result, however, it deviates from the N-body result by more than one percent already for $k\agt 0.05 \;\mathrm{h\, Mpc^{-1}}$. The reason for the poor performance of  eq.~\re{PTRGRSnA}, is the absence of any term accounting for the so-called FoG effect. It can be included  in a purely phenomenological way by multiplying the full eq.~\re{PTRGRSnA} by a scale-dependent function, which in this case we take of the exponential form, $e^{-A k^2}$.  Notice that, in general, the FoG modelling depends on the line of sight direction, $\mu$, however, since we are considering only the monopole, we just restrict ourselves to a purely phenomenological, scale dependent, function.
In this case the fit on the full PS clearly improves, as illustrated by the red solid lines in Fig.~\ref{PratioRS}.

\begin{figure}
\includegraphics[width=0.8\textwidth,right]{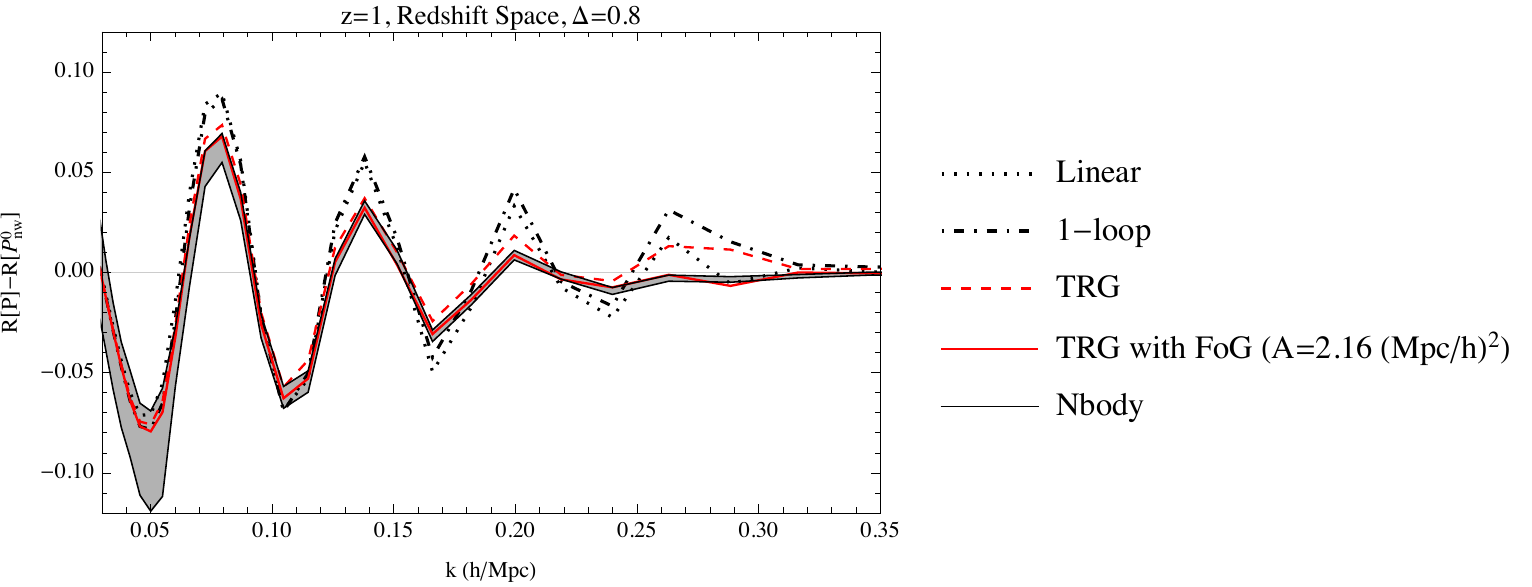}
\includegraphics[width=0.8\textwidth,right]{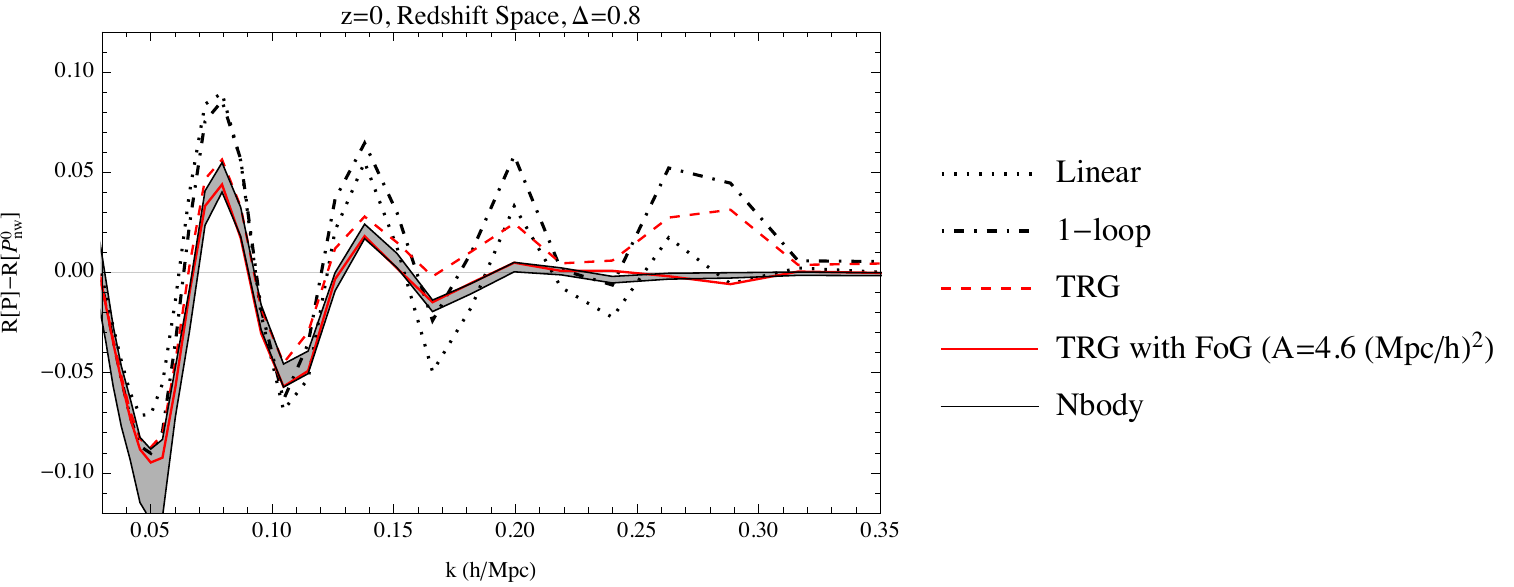}
\caption{The extractor $R[P](k;\Delta)$ applied to the redshift space monopole PS computed from N-body simulations (grey area), from linear theory (blue, dashed), 1-loop SPT (black, dash-dotted), TRG (red), and ``TRG+FoG corrections'' (orange), as described in the text, at redshift $z=1$ and at $z=0$. The grey area for the N-body results corresponds to the error computed using eq.~\re{DRdisc} and using the  1-$\sigma$ error on the PS for each bin. For visualisation purposes, the same quantity $R[P^{0,nw}](k)$, where $P^{0,nw}$ is the smooth component of the linear PS, has been subtracted from all the different $R[P](k)$. }
\label{newfig6RS2}
\end{figure}

\begin{figure}
\includegraphics[width=0.8\textwidth,right]{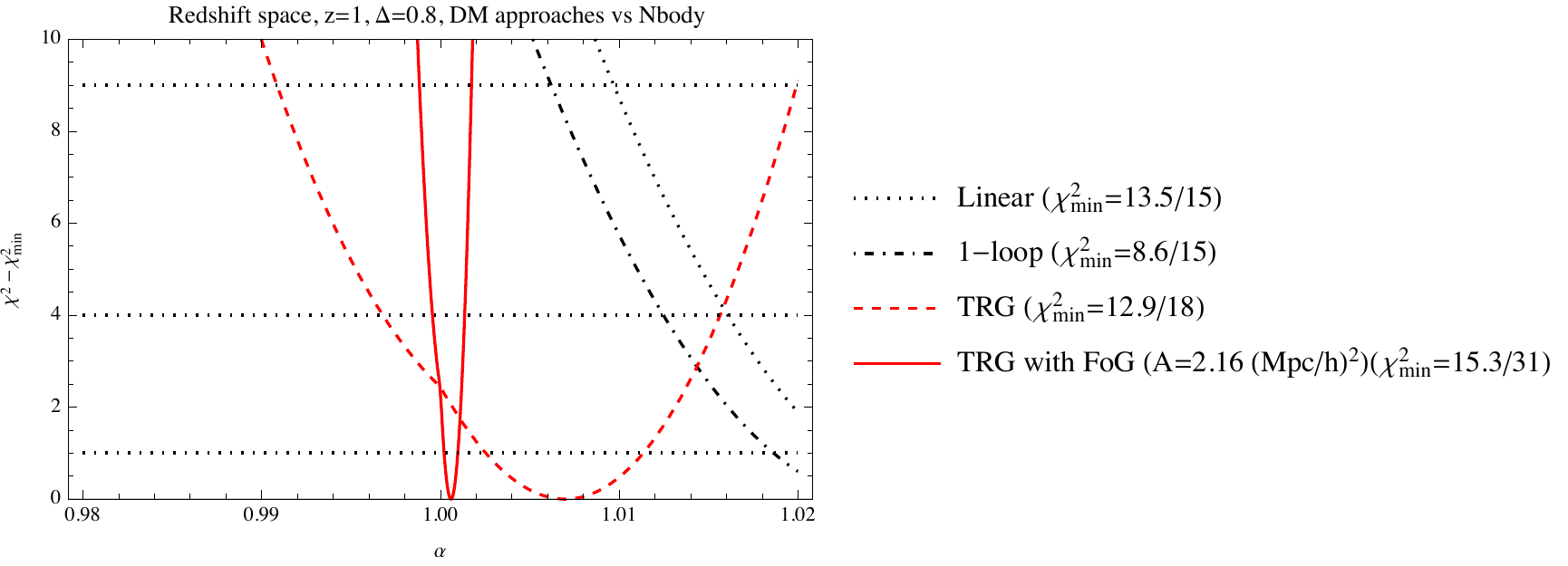}
\includegraphics[width=0.8\textwidth,right]{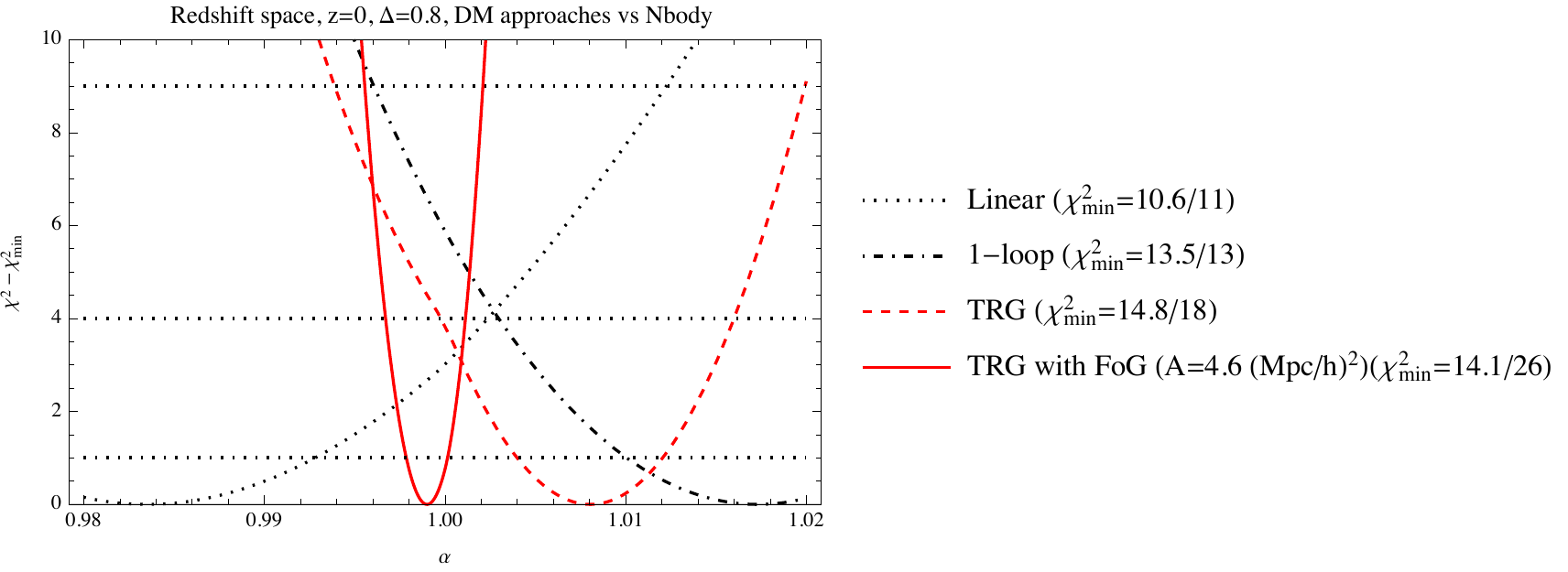}
\caption{ $\chi^2$ as a function of the shift parameter $\alpha$ for different DM approaches vs N-body in redshift space.}
\label{chivsNB3}
\end{figure}

As far as BAO's are concerned, the  FoG-corrected TRG gives a very good fit to  the N-body $R[P]$ over all the relevant range of scales, while the TRG without FoG correction clearly performs better than linear theory and 1-loop SPT, at least for the lower peaks, see Fig.~\ref{newfig6RS2}. To be more quantitative, we define a new likelihood function, analogous  to eq.~\re{chi2}, in which the role of $P_{data}$ is taken by the redshift space (monopole) PS from N-body simulations. As we see from Fig.~\ref{chivsNB3}, the inclusion of the FoG correction on the TRG result reproduces the N-body BAO scale at the $0.1\%$   level, in line with its real space counterparts (see Fig.~\re{chivsNB}), while without FoG correction it still performs at better than the percent level.

\subsection{Halo bias}
\label{sec:halobias}
We now study the effect of halo bias on the function $R[P]$. We focus our discussion on halos of mass $M > 10^{13} \, M_{\odot}$ identified in the N-body simulations  described in \ref{Simul}, where we also plot the bias, that is the ratio between the halo and DM PS's, for the complete halo catalog and for its partition in different mass bins, see Fig.~\ref{fig:bias}. We also list, in 
Tab.~\ref{table:Mass_bins}, best fit values for a model bias function given by $(b_0+b_1 k^2)^2$. As the $R[P]$ operation is insensitive to a constant bias, we consider in this section, as we did in Sect.~\ref{sec:themodel}, a single parameter model for the halo, again of the exponential form, $e^{-A k^2}$. More refined models can of course be tested but as we will see, this one already provides a very good fit.
\begin{figure}
\includegraphics[width=0.8\textwidth,right]{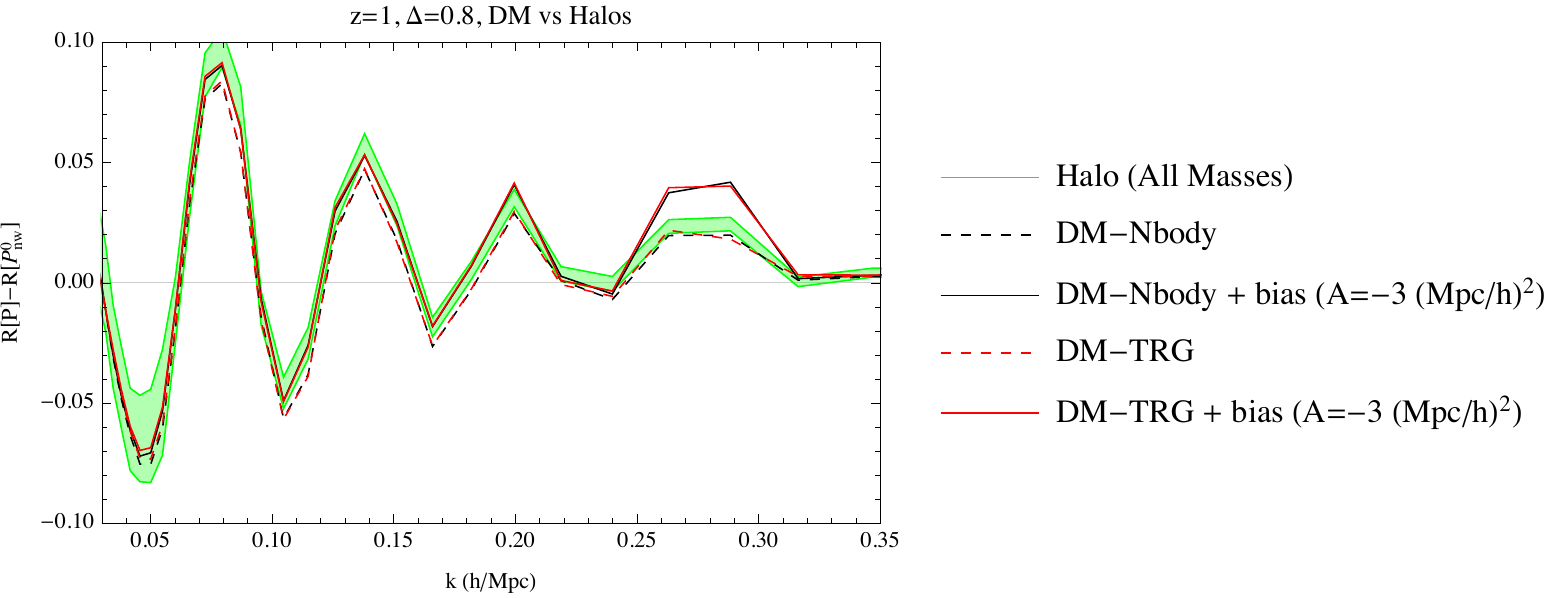}
\includegraphics[width=0.8\textwidth,right]{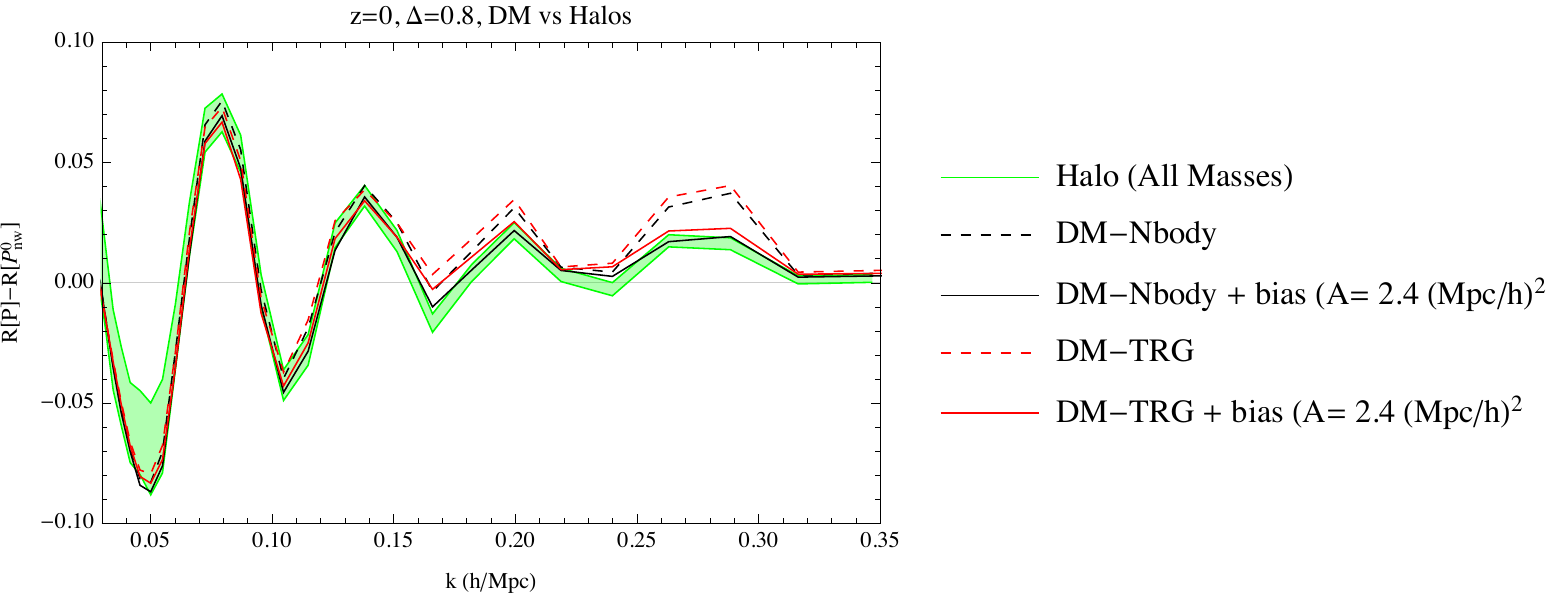}
\caption{The extractor $R[P](k;\Delta)$ applied to 
the PS of all halos with mass $>10^{13} \, M_\odot$ (green band). The black dashd lines are obtained from the DM PS from N-body simulations, while the red dashed lines are obtained from the TRG. The corresponding solid lines are obtained by multiplying the PS by $e^{-A k^2}$, with $A$ fitted from the smooth component.
}
\label{Rbias}
\end{figure}

\begin{figure}
\includegraphics[width=0.8\textwidth,right]{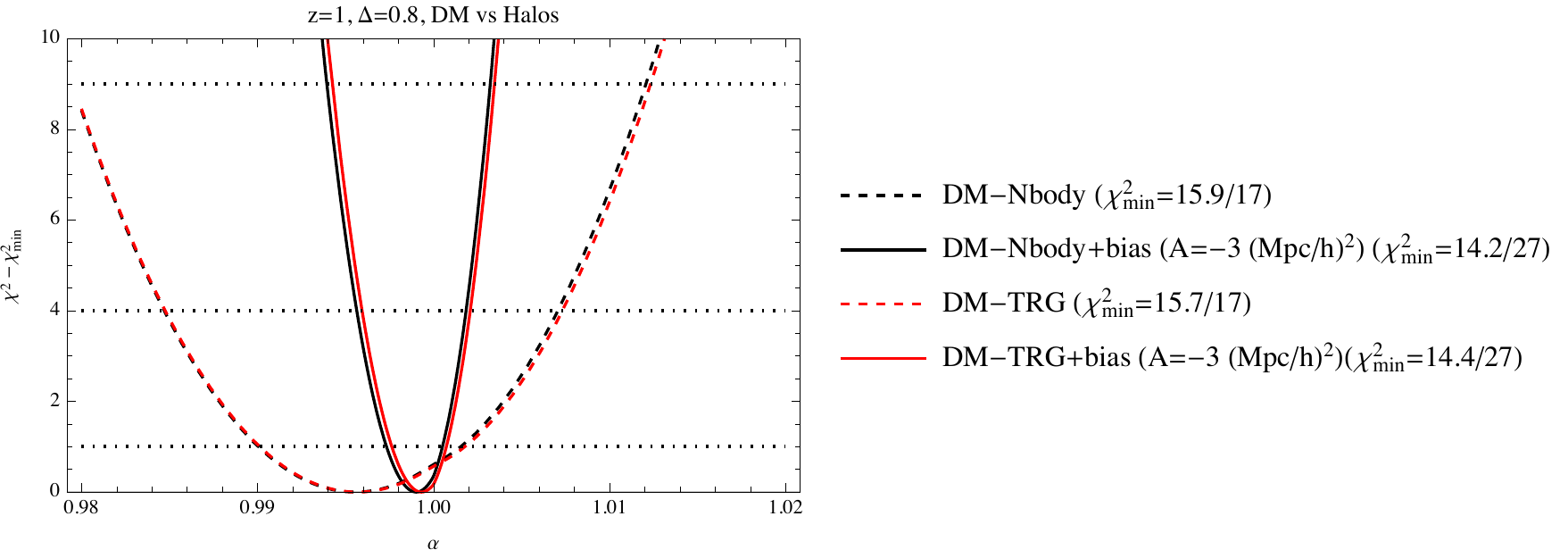}
\includegraphics[width=0.8\textwidth,right]{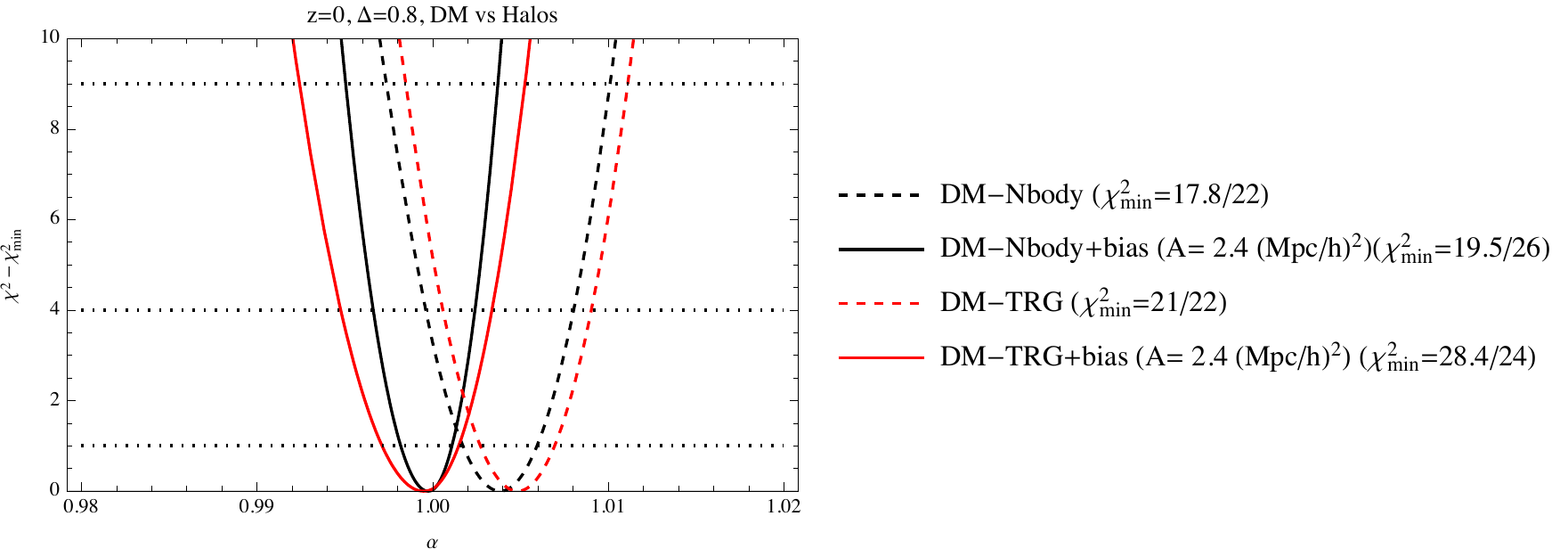}
\caption{$\chi^2$ as a function of the shift parameter $\alpha$ for DM (from N-body and TRG) vs. halos with $M >10^{13} \, M_\odot$. }
\label{fig:chi2halo}
\end{figure} 
The different PS give the $R[P]$'s plotted in Fig.~\ref{Rbias}. The green band is obtained from the total halo PS, while the DM PS from N-body simulations is given by the black lines: the dashed ones are obtained from the unbiased PS, while the  solid ones are from the one multiplied by $e^{-A k^2}$, with $A$ fit from the total halo PS bias. The red lines are obtained by the TRG approximation, again unbiased (dashed) and biased (solid).

Then, taking the halo PS (with the corresponding errors) as $P_{data}$ in eq.~\re{chi2},  we study the shift in the BAO scale derived by modeling it using the different procedures.  The results are given in Fig.~\ref{fig:chi2halo}. The dashed curves show that the effect of bias is $O(0.5 \%)$ both at $z=0$ and at $z=1$, although opposite in sign. Notice, again, that what counts here is the scale dependence of the bias, which is in magnitude roughly the same at the two different redshifts. The bias can be entirely taken into account by our simple exponential function, and  the TRG results practically coincide with the N-body ones. Similar results are obtained in redshift space, as one can infer also by looking at Fig.~\ref{BaoExtr}.

\section{Conclusions}
\label{sec:conclusions}
In this work we studied and furhter developed the estimator $R \left[ P \right]$ introduced in ref. \cite{Noda:2017tfh} to extract the BAO scale from a power spectrum $P$. We showed that the estimator is extremely robust against various physical effects. The estimator is mostly sensitive to the nonlinear evolution of the DM field, which causes a rescaling of the BAO scale of $\agt 1\%$.  Redshift space distortions beyond the (improved) Kaiser approximation and halo bias affect the $R \left[ P \right]$ {extraction} only at the sub-percent level at all redshifts, as indicated, respectively, by  Figures \ref{chivsNB2}  and \ref{fig:chi2halo}, and can be taken into account by simple fitting functions, such as the $e^{-A k^2}$ one considered in this paper.
  Once the parameter $A$ is fitted to the {\em broadband shape} of the PS measured from real data, the extraction of the BAO scale from the $R[P]$ projector is a parameter-free procedure. 

 We studied how the nonlinear evolution of the DM field can be taken into account in SPT and in the TRG approach described in \cite{Noda:2017tfh}, which improves over 1-loop SPT by including the effect of large scale motions and by correcting the short-scale behavior via the addition of effective coefficients, measured from numerical simulations.
As seen in Figure \ref{Pratio}, in comparison to 1-loop SPT, the short-scale corrections significantly improve the {reproduction} of the broadband shape of the PS, and are therefore relevant when the latter matters, as, for instance, in studying the effect of massive neutrinos \cite{LMPR09, Senatore:2017hyk, Upadhye:2017hdl}~\footnote{Further improvement is required to accurately reproduce the effects of the  FoG in redshift space. Improving over the existing phenomenological approaches requires  going beyond the single stream approximation.}. However, their impact on the BAO {scale} is greatly alleviated. This proves that the extraction of the BAO scale is to a large extent independent of UV effects, so that expensive simulations (required to either solve the dark matter dynamics, or to compute effective coefficients) are not needed. The main improvement of the TRG method  over  1-loop SPT as far as BAO are concerned is in the resummation of the effect of large scale  modes, which provides an accurate accounting of the oscillating structure of $R[P]$ over all scales and at all redshifts, see Fig.~\ref{newfig6}. The exponential damping term accounting for such effects in the TRG approach can be immediately computed with a 1-dimensional integral of the linear PS. 

We compared our extraction procedure to the standard approach to the BAO scale measurements,  which involves a multi-parameter fit to the full PS and makes use of fitting functions for the smooth linear PS. Beside having considerably fewer parameters (or not at all), our approach, when applied to the halo catalog considered in this paper, reduces the error on the extracted scale by a factor $\sim 4.7$ ($\sim 3.6$) at redshift $z=0$ ($z=1$), as shown in Figure \ref{BaoExtr}. 

In principle, the procedure can be applied to the {\it reconstructed} PS, obtained after the long range displacements have been undone \cite{Eisenstein:2006nk,Padmanabhan:2008dd,Noh:2009bb,Tassev:2012hu,Sherwin:2012nh,White:2015eaa}: therefore it is not an alternative to reconstruction, but rather, it provides a parameter independent procedure to extract BAO information from reconstructed data. We plan to investigate this possibility in a forthcoming work.

In summary, the BAO extracting procedure outlined in Sect.~\ref{sec:R[P]}, and the simple model introduced in Sect.~\ref{sec:themodel} allow to compute the BAO scale at sub-percent accuracy, at all redshifts, including bias and redshift space distortions effects. This procedure is extremely fast, and no more computationally expensive than 1-loop (SPT). 
Moreover, contrary to the standard procedure, it does not require a multi-parameter fit of the data to account for the broadband shape of the PS, which is largely filtered out by the extractor.

\section*{Acknowledgments}
E. Noda and M. Pietroni acknowledge support from the European Union Horizon 2020 research and innovation programme under the Marie Sklodowska-Curie grant agreements Invisible- sPlus RISE No. 690575, Elusives ITN No. 674896 and Invisibles ITN No. 289442.
The work of M. Peloso was supported in part by DOE grant de-sc0011842 at the University of Minnesota.
T. Nishimichi acknowledges financial support from Japan Society for the Promotion of Science (JSPS) KAKENHI Grant Number 17K14273 and Japan Science and Technology Agency (JST) CREST Grant Number JPMJCR1414.
Numerical calculations for the present work have been carried out on Cray XC30 at 
Center for Computational Astrophysics, CfCA, of National Astronomical Observatory of Japan. 
\appendix
\section{TRG}
\label{appendicetrg}

Here we summarize the Time Renormalization Group (TRG) system of equations for the PS derived in   \cite{Noda:2017tfh} and used in this work. The PS comprises of two parts: a smooth broadband (``no-wiggle'') component $P^{ nw}$, and a smaller (``wiggle'') part $P^{ w}$  that contains the BAO oscillations. We consider two separate equations for the two components. As a starting point we separate the linear power spectrum in a smooth plus an oscillatory component, $P^0 = P^{0,{ nw}} +  P^{0,{w}}$. To achieve this, we notice that the oscillatory part is of the form $P^{0,{ w}} = {\tilde P} \left( k \right) \, \sin \left( k \, r_{bao} \right)$, where $ {\tilde P} \left( k \right)$  is a smooth 
function, and the scale $r_{bao}$ can be estimated, for any given cosmology, through eq.  (6) of \cite{Komatsu:2008hk}. We construct $P^{0,{ nw}}$ by evaluating the PS at the nodes of $\sin \left( k \, r_{bao} \right)$ and by interpolating in a smooth way between these values. 

We then use $P^{0,{ nw}}$ to compute the one-loop SPT correction $\Delta P^{1-loop,nw}$.  

SPT fails to accurately reproduce the PS at scales $k \ga 0.1 \, h \, {\rm Mpc}^{-1}$. This is due both to UV effects that go beyond the single stream approximation, and to IR effects, such as large scales bulk flow. As discussed in  \cite{Noda:2017tfh}, UV effects mostly impact the smooth component  $P^{nw}$, and can be accounted for by including additional terms in the TRG evolution equation for the PS. These terms can be evaluated through N-body simulations \cite{Pietroni:2011iz,Manzotti:2014loa,Noda:2017tfh}. While this correction considerably improves the PS at  $k \ga 0.1 \,  {\rm h\;Mpc}^{-1}$, it does not affect the BAO oscillation scale extracted from $R[P]$ \cite{Noda:2017tfh}. We demonstrate this in Section \ref{sec:themodel}, where we show that the scale extracted from the halo catalog from our numerical simulations is in excellent agreement with that extracted from a theoretical PS obtained without this UV terms. 

On the contrary, IR bulk flows decease the coherency of the BAO oscillations, and need to be included in order to accurately reproduce the oscillatory behavior of $P^{w}$. These effects can be resummed in an exponential (scale-dependent) suppression of the oscillatory term,  given by the last term of eq. (\ref{PtrgRS}). While this result case derived in \cite{Noda:2017tfh} on the basis on the consistency relations between the  power spectrum and the squeezed bispectrum (the soft mode of which is a IR bulk flow) \cite{Peloso:2013zw}, in \ref{appendiceIR} we provide an alternative proof that can be more immediately extended to redshift space.

\section{IR resummation}
\label{appendiceIR}
In this appendix we study how IR bulk flows resum to the exponential term in eq.~(\ref{PtrgRS}) of the main text.  This effect of the IR modes on the oscillatory part of the PS was derived in real space in  \cite{Noda:2017tfh}. We rederive this result using a Lagrangian formulation, that can be easily extended to redshift space. 

We consider a large set of particles, and we split their velocities in long (``l'') and short (``sh'') wavelength components, treating the former in linear theory (the splitting can be done, for instance, by computing the average velocity of the particles contained in a ``large'' volume, and splitting the velocity of each particle in that volume as the average plus a residual). We denote as $f$ the phase space distribution of the particles and as $f^{\rm sh}$ the distribution that the particles would have  if their position was not changed by the long wavelength velocity. The two distributions are related by 
\beq
f \left( \bx,\bp^l,\bp^{sh},\eta \right)  =\int d^3 y \,f^{sh} \left( \by,\bp^l,\bp^{sh},\eta \right)  \delta_D \left( \by-\bx +\frac{\bp^l}{a m {\cal H} f } \right)\,, 
\label{f-fsh}
\eeq
where $\bx$ and $\by$ are spatial coordinates, $\delta_D$ the Dirac-delta function, and where the comoving momentum of the particles is related to their velocity by the non relativistic expression $\bv = \frac{\bp}{a \, m}$ (where $a$ is the scale factor and $m$ the mass of the particles). Integrating the distribution function over the momentum gives the particle number density
\beq
n \left( \bx,\eta \right)=\bar n\left(1+\delta(\bx,\eta)\right)=\int d^3 p^ld^3 p^{sh} f(\bx,\bp^l,\bp^{sh},\eta)\,,
\label{n-def}
\eeq
where $\bar n$ is the average number density, and $\delta \left( \bx \right)$ the density contrast. By Fourier transfroming this relation, and using eq. (\ref{f-fsh}), we can write 
\beqra
(2\pi)^3\delta_D(\bk) + \delta(\bk,\eta) &=&  {\bar n}^{-1} \, \int d^3 y d^3 p^l\,d^3 p^{sh} e^{-i\bk\cdot\by}e^{-i\frac{\bk\cdot\bp^l}{am{\cal H} f}}f^{sh}(\by,\bp^l,\bp^{sh},\eta) \nonumber\\
&=& 
\int d^3 y \,e^{-i\bk\cdot\by}\left(1+\delta_{sh}(\by,\eta)\right) e^{-i\frac{\bk\cdot\bv^l(\by,\eta)}{{\cal H}f}+\cdots}\,,
\label{shortdress}
\eeqra
where the second line has been obtained by expanding the exponent in the first line around $\bp^l = 0$, where 
$\delta_{sh}$ is defined from $f^{sh}$ as in eq. (\ref{n-def}), where $bv^l$ is the first moment 
\beqra
\left(1+\delta_{sh}(\by,\eta)\right) \bv^l(\by,\eta)  =\bar n^{-1} \int d^3 p^l\,d^3 p^{sh} \,\frac{\bp^l}{a m} \,f^{sh}(\by,\bp^l,\bp^{sh},\eta)\,,
\eeqra
and where the ellipsis includes the higher moments that we disregard. From (\ref{shortdress}) we obtain the effect of the long wavelength modes on the PS 
\beqra
&& \!\!\!\!\!\!\!\!  \!\!\!\!\!\!\!\! 
P(k) = \int d^3 r\,e^{-i\bk\cdot\br} \left\langle (1+\delta_{sh}(\br/2,\eta))  (1+\delta_{sh}(-\br/2,\eta))  e^{-i\frac{\bk\cdot 
\left[   \bv^l(r/2)-\bv^l(-r/2) \right]}{{\cal H} f}} \right\rangle \nonumber\\ 
\!\!\!\!\!\!\!\! && \simeq  \int d^3 r\,e^{-i\bk\cdot\br}   \left\langle (1+\delta_{sh}(\br/2,\eta))  (1+\delta_{sh}(-\br/2,\eta)) \right\rangle e^{-\int^\Lambda \frac{d^3 q}{(2 \pi)^3} \frac{(\bk\cdot\bq)^2}{q^4} (1-\cos(\bq\cdot\br))P^0(q)} \,. \nonumber\\ 
\label{PIR}
\eeqra
To obtain the second line, we have treated $\bv^l$ as linear, gaussian, and uncorrelated with $\delta_{sh}$. The cut-off  $\Lambda$ is introduced as a reminder that only IR modes contribute to  $\bv^l$. ~\footnote{The dependence of the results on the cut-off $\Lambda(k)$ has been discussed in \cite{Noda:2017tfh}, where it has been showed that, as long as a $\Lambda$CDM linear PS is considered, omitting the cut-off still provides a numerically accurate description of the IR resummation. For this reason we also set $\Lambda(k)=\infty$ in our computations, and we omit the $k$-dependence from the argument of $\Xi$.} 

If we consider the oscillating component of the PS, namely, 
\beq
\!\!\!\!\!\!\!\!\!\!\!\!\!\!\! P^w_{sh}(k) \equiv \int d^3 r e^{- i\bk\cdot\br}  \left\langle (1+\delta^w_{sh}(\br/2,\eta))  (1+\delta^w_{sh}(-\br/2,\eta)) \right\rangle = {\tilde P(k)} \sin(k\,r_{bao}) \,, 
\label{Pw}
\eeq
eq.~\re{PIR} gives a simple result.
We insert this expression into (\ref{PIR}), and we expand the exponential term 
\beqra
&&\!\!\!\!\!\!\!\!\!\!\!\!\!\!\!\!\!\!\!\!\!\!\!\! P^w(k)= \sum_{n=0}^\infty \frac{(-1)^n}{n!}\int d^3 r \,e^{-i\bk\cdot\br} \int \frac{d^3 p}{(2 \pi)^3} e^{i\bp\cdot\br}   \tilde P(p) \sin(p\,r_{bao}) \times\nonumber\\
&&\!\!\!\!\!\!\!\!\!\!\!\!\!\!\!\!\!\!\!\!\!\!\!\!\!\!\!\!\!\! \!\!\ \int^\Lambda \frac{d^3 q_1}{(2\pi)^3}\frac{(\bk\cdot\bq_1)^2}{q_1^4} P^0(q_1) \left(1-\cos(\bq_1\cdot\br)\right)\cdots \int^\Lambda \frac{d^3 q_n}{(2\pi)^3}\frac{(\bk\cdot\bq_n)^2}{q_n^4} P^0(q_n) \left(1-\cos(\bq_n\cdot\br)\right)\,.\nonumber\\
\label{PwIR}
\eeqra

Let us focus on the $n=1$ term. Integrating over $r$ introduces a series of $\delta_D$ functions, that can be used in the integral over $p$, to give 
\beqra
&&\!\!\!\!\!\!\!\!\!\!\!\!\!\!\!\!\!\!\!\!\!\!\!\! \!\!\!\!\!\!\!\! 
- \int \frac{d^3 q_1}{\left( 2 \pi \right)^3} P^0 \left( q_1 \right) \, \frac{\left( \bk \cdot \bq_1 \right)^2}{q_1^4} \left\{ 
{\tilde P} \left( k \right) \sin \left( k \, r_{\rm BAO} \right) 
- \sum_{r_1=\pm} \frac{1}{2} {\tilde P} \left( \vert \bk + r_1 \bq_1 \vert \right) \sin \left( \vert \bk+ r_1 \bq_1  \vert \, r_{\rm BAO} \right)  \right\} \nonumber\\ 
&&\!\!\!\!\!\!\!\!\!\!\!\!\!\!\!\!\!\!\!\!\!\!\!\! \simeq - 
\int \frac{d^3 q_1}{\left( 2 \pi \right)^3} P^0 \left( q_1 \right) \, \frac{\left( \bk \cdot \bq_1 \right)^2}{q_1^4} \, 
{\tilde P} \left( k \right) \sin \left( k \, r_{\rm BAO} \right) \left[ 1 - \cos \left( \bq_1 \cdot {\hat \bk} \, r_{\rm BAO} \right) \right] \;, 
\label{IR-n=1}
\eeqra
where the second line has been obtained by expanding the term in parenthesis in the first line in the limit of $q_1 \ll k$ (which is appropriate, as we are considering the effect of IR modes). A direct inspection of the $n=2,3,\dots$ terms show that the $\int d^3 r d^3 p$ integration produces an expression that can be simplified analogously to (\ref{IR-n=1}) (one can expand recursively over the $\bq_i$ momenta). This leads to 
\beqra
P^w(k) \simeq P_{sh}^w \left( k \right) \sum_{n=0}^\infty \frac{(-1)^n}{n!} \left[ k^2 \,  \Xi \left( r_{\rm bao} \right) \right]^n 
= {\rm e}^{-k^2 \,  \Xi \left( r_{\rm bao} \right) } \,  P_{sh}^w \left( k \right) \;, 
\label{Pw-real}
\eeqra
where 
\beqra
&& \Xi(r) \equiv \frac{1}{k^2}  \int^\Lambda \frac{d^3 q}{(2\pi)^3}\frac{(\bk\cdot\bq)^2}{q_1^4} P^0(q) \left(1-\cos(\bq\cdot \hat\bk\,r_{bao})\right) \nonumber\\
&& \quad \;\;\;= \frac{1}{6 \pi^2}\int^\Lambda dq P^0(q) (1-j_0(q r)+2j_2(q r))\,. 
\eeqra

This computation can be readily extended to redshift space. In this case, the relation (\ref{shortdress}) is modified into 
\beqra
\!\!\!\!\!\!\!\!\!\!\!\!\!\!\!\!\!\!\!\!&& 
(2\pi)^3\delta_D(\bk) + \delta_s(\bk)= 
 \int d^3 y\, e^{-i\bk\cdot\by}  \,e^{-i\bk\cdot\by}\left(1+\delta_{s,sh}(\by)\right) e^{-i\frac{\bk\cdot\bv^l(\by)}{{\cal H}f} -i\frac{k_z v^l_z(\by)}{{\cal H} }+\cdots }\,,
\eeqra
where $\delta_s$ is the density contrast in redshift space, and the $z-$direction is orient along the line of sight. Starting from this expression, and repeating the same steps done to obtain (\ref{Pw-real}) now leads to 
\beq
P^s_w(k,\mu)=e^{-k^2\Xi(r,\mu)} P^{s,sh}_w(k)\,,
\label{Pw11}
\eeq
where 
\beq
\!\!\!\!\!\!\!\!\! \!\!\!\!\!\!\!\!\!  \!\!\!\!\!\!\!\!\!   \Xi(r,\mu)\equiv  (1+f \mu^2(2+f))\Xi(r) + f^2\mu^2(\mu^2-1) \frac{1}{2 \pi^2} \int^\Lambda dq P^0(q) j_2(q r) \,,
\label{24}
\eeq
and $P^{s,sh}_w(k)$ is the oscillatory part of the PS obtained from the $\delta_{s,sh}(\by)$ field.

\section{N-body simulations}
\label{Simul}
We employ a set of N-body simulations created using the public GADGET-2 code  \cite{Springel:2005mi}. The data consisted in cubic boxes of comoving side length $L_{box}= 2048$  $h^{-1}$ Mpc and $2048^3$ particles . The cosmological parameters follow those of the Planck 2015 results  \cite{Ade:2015xua}, and are reported in Table \ref{table:cosmology}.

The initial condition were created at redshift z=29.4 using the generator developed in \cite{Nishimichi:2008ry} and parallelized in  \cite{Valageas:2010yw}. The particles were displaced from a uniform grid using second order Lagrangian perturbation theory (2LPT) \cite{Scoccimarro:1997gr,Crocce:2006ve}. The initial redshift is somewhat lower than the values commonly used in literature, but it has been in shown that a higher starting redshift does not improve the result \cite{Taruya:2012ut}.

 We employed the method by Angulo and Pontzen \cite{Angulo:2016hjd} to suppress cosmic variance. It consists in running two simulations which have initial conditions with the same magnitude, but opposite phase and then taking the average of the PS obtained from them. In this work we used the data collected at redshift $z=1.10911$ (in the rest of the paper just indicated with $z=1$) and $z=0$. 
In \ref{COVNB} we discuss in detail how the diagonal elements of the covariance matrix from our simulations have been measured, and compute the nondiagonal ones in perturbation theory, showing that their effects on our analyses is negligible.

From the data we constructed a halo catalog using the ROCKSTAR halo finder \cite{Behroozi:2011ju} which is based on an adaptive algorithm for refining friend-of-friend groups of particles looking at their six-dimensional phase space distribution. ROCKSTAR also keeps track of the time evolution to improve the consistency of the hierarchy of substructures throughout the evolution. In our work we kept halos more massive than $10^{13}$ solar masses. 
Using the Cloud-in-Cell interpolation we constructed the density contrast field for the Dark Matter and Dark Matter Halos, both in real space and in redshift space (using the distant observer approximation). From these we then calculate the relevant PS after moving to Fourier space, via FFT.

\begin{figure}
\centering{ 
\includegraphics[width=0.45\textwidth,clip]{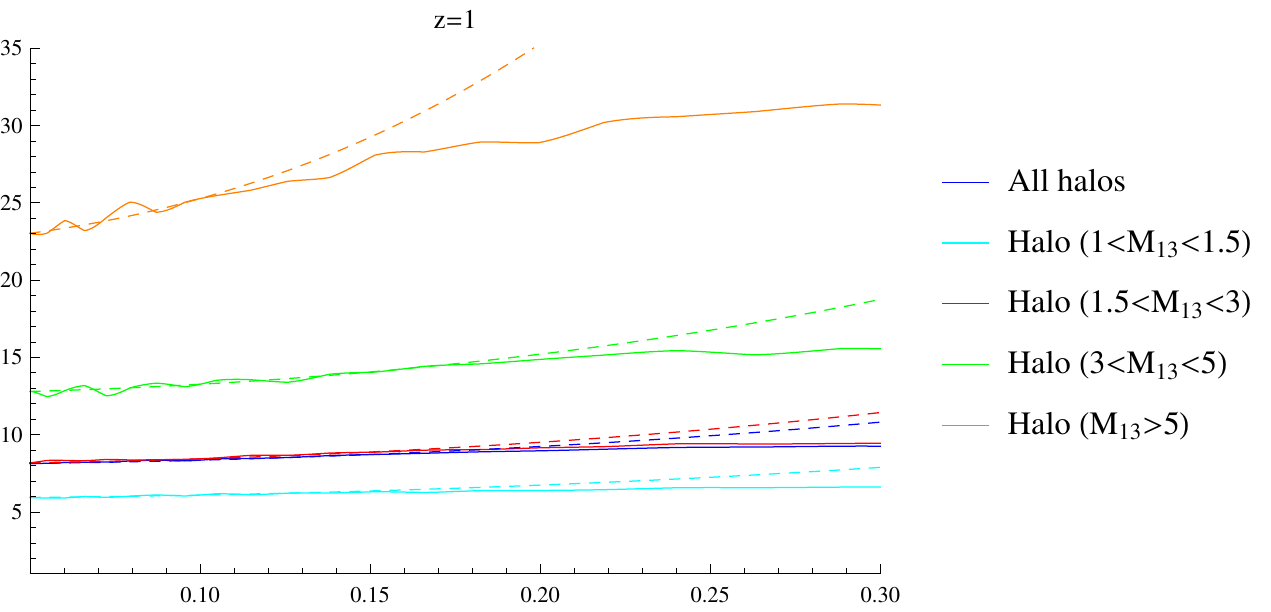}
\includegraphics[width=0.45\textwidth,clip]{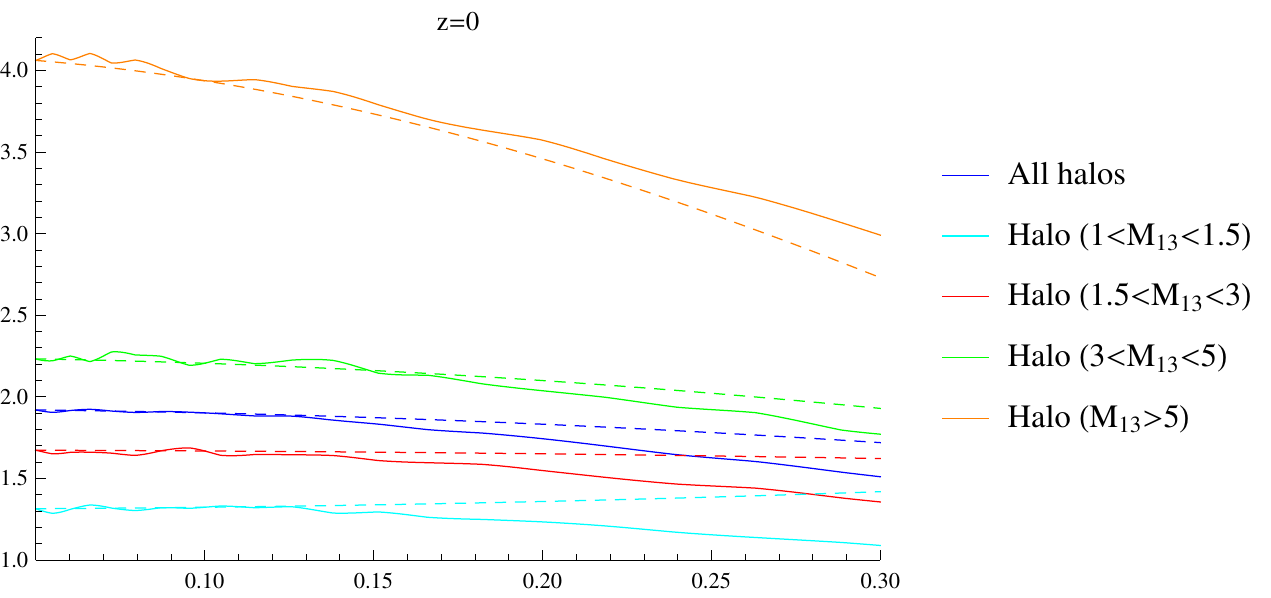}
}
\caption{Bias for each halo bin and for the total halos of $M >10^{13} \, M_\odot$ in our simulations. The solid lines are the ratio between the halo and the matter power spectra. The dashed lines is the scale dependent bias $\left( b_0 + b_1 \, k^2 \right)^2$, with the coefficients chosen so to match the ratio at $k=0.05 \, h / {\rm Mpc}$ and at $k=0.1 \, h / {\rm Mpc}$. As we study in Section \ref{sec:halobias}, obtaining a good bias function at these scales allows for a very good {extraction} of the BAO oscillation pattern.
}
\label{fig:bias}
\end{figure}

At this point we can estimate the effect of bias on the nonlinear PS. In figure \ref{fig:bias} we plot the ratio between the halo density PS and the underlying DM PS at redshift $z=0$ and $z=1$, divided in mass bins. The number of halos in each bin is shown in Table \ref{table:Mass_bins}.

We use $ P_{halo}(k) = (b_0 + b_1 k^2 )^2 P_{DM}(k) $  as the scale-dependent model for the halo bias. Although it does not fit well the data at all scales, we verified that in order to extract the BAO it is sufficient to fit just the largest scales. To this end, we obtain the coefficients $b_0$ and $b_1$ by imposing that the bias function reproduces the ratio between the halo and matter PS at $k=0.05\; \mathrm {h\;Mpc^{-1}}$ and at $k=0.1 \;\mathrm {h\;Mpc^{-1}}$. The resulting coefficients are listed in Table \ref{table:Mass_bins}, and the performance of the fitting functions are of the same quality than those shown in Figure 
\ref{fig:bias}, which have been obtained by using an exponential function $e^{-A k^2}$.

\begin{table}
\centering
\begin{tabular}{ |c c c c c c |}
 \hline
$ \Omega_{m} $ & $\Omega_{\Lambda} $ & $\Omega_b$ &  h & $\ln(10^{10} A_s) $ & $n_s$ \\
 \hline
 0.3156 & 0.6844 & 0.0491 & 0.6727 & 3.094 & 0.9645\\
 \hline
\end{tabular}
\caption{Cosmological parameters}
 \label{table:cosmology}
 \end{table}

\begin{table}
\centering
\begin{tabular}{ |c |c c c |c c c |}
 \hline
  & & $z=0$ & & & $z=1$  & \\
  \hline
  Mass range & $N_{halos} $ & $b_0$ & $b_1$  & $N_{halos} $& $b_0$ & $ b_1$\\ 
   \hline
 $M_{halo} \geq 10^{13} M_{\odot}$  (All masses) & 3890690 & 1.387 & -0.848 & 1763542 &2.839 & 4.976\\ 
 $1.0 \times 10^{13} M_{\odot} \leq M_{halo} \leq 1.5 \times 10^{13} M_{\odot}$ & 1338222 & 1.145 &  0.510 & 796154 & 2.425	&	4.242\\ 
 $1.5 \times 10^{13} M_{\odot} \leq M_{halo} \leq 3.0 \times 10^{13} M_{\odot}$ & 1370335 & 1.293 &  -0.226 & 665216 & 2.844& 5.953\\ 
 $3.0 \times 10^{13} M_{\odot} \leq M_{halo} \leq 5.0 \times 10^{13} M_{\odot}$ & 547274 & 1.497 &  -1.204& 191029 & 3.555 & 8.608\\ 
  $M_{halo} \geq 5.0 \times 10^{13} M_{\odot}$ & 634859 & 2.026 & -4.150 & 111144 &4.722 &30.45\\ 
 \hline
\end{tabular}
\caption{The table shows the number of halos in our Nbody simulation for $z=0$ and $z=1$, averaged over realizations and divided in different bins according to their mass. The other columns refer to the parameters for the halo PS, where we used $P_{halo}(k) = (b_0 + b_1 k^2)^2 P_{Nbody}(k)$ as the scale dependence bias funtion (the coefficient $b_1$ is given in units of $h^{-2} \, {\rm Mpc}^2$).}
 \label{table:Mass_bins}
\end{table}

\section{Covariance matrix of the Power Spectrum from N-body simulations in the Angulo-Pontzen method.}
\label{COVNB}
In the analyses presented in this paper, we ignore the off-diagonal components of the covariance matrix for the PS, while the diagonal part is estimated from the scatter of $|\delta_{\vec{k}}|^2$ for $\vec{k}$'s in each of the k bins. In this appendix we discuss the covariance matrix for the PS obtained in the Angulo-Pontzen method \cite{Angulo:2016hjd} and compute the effects of its nondiagonal entries using PT and with a new set of simulations.\\
We start from the definition of covariance matrix for the PS. Given a realization, we can define a PS estimator in the momentum bin $k_m$ as \cite{Scoccimarro:1999kp} (note that our $(2\pi)^3$ convention is different from that paper)
\beq
\hat P_m = \frac{1}{V} \int_{k_m} \frac{d^3 k}{V_s(k_m)} \delta_\bk\delta_{-\bk}\,,
\eeq
where the integral is made over a momentum shell centered at $k_m$, $V_s(k_m) = 4\pi k_m^2 \delta k_m$, with $\delta k_m$ the width of the $m$'th bin,  and $V$ is the volume of the simulation/survey. Moreover, we have $(2 \pi)^3\delta_D(0)=V$.\\
Then, by considering an ensemble of realizations, we define the covariance matrix as 
\beqra
\cov^P_{mn}&=&\langle \hat P_m\hat P_n\rangle - \langle \hat P_m\rangle\langle\hat P_n\rangle\nonumber\\
&=&\frac{(2\pi)^3}{V}\left[ \frac{2 P_m^2}{V_s(k_m)}\delta_{mn} + \frac{1}{(2\pi)^3} \bar T_{mn}\right]\,,\nonumber\\
&=&\langle \delta P_m\,\delta P_n\rangle\,,
\label{covdef}
\eeqra
where $P_m =\langle \hat P_m\rangle $ is the average of the PS among the realizations and  $\bar T_{mn}$ is the bin-averaged trispectrum
\beqra
&&\!\!\!\!\!\!\!\!\!\!\!\!\bar T_{mn}\equiv \int_{k_m} \frac{d^3 k}{V_s(k_m)}\int_{k_n} \frac{d^3 k'}{V_s(k_n)} T(\bk,-\bk,\bk',-\bk')\,,\nonumber\\
&&\!\!\!\!\!\!\!\!=\frac{1}{2}\frac{1}{ k_m^2 k_n^2 \delta k_m \delta k_n} \int_{k_m}^{k_m+\delta k_m} k^2 dk \int_{k_n}^{k_n+\delta k_n} k'^2 dk' \int_{-1}^{1} d\cos\theta \,\tilde T(k,k',\cos\theta)\,,
\eeqra 
where in the last line we have used rotational invariance.
At the third line of \re{covdef} we have defined  $\delta P_m= \hat P_m- P_m$.\\
In the Angulo-Pontzen method \cite{Angulo:2016hjd}, hereafter ``AP method", the linear density field (the initial condition for the simulation) is not sampled from a gaussian distribution but is set as follows,
\beq
\delta^{(0)}_\bk= \sqrt{V P^0(k)} \,e^{i\theta_\bk}\,,
\label{apic}
\eeq
where $\theta_\bk=-\theta_{-\bk}$ is drawn with uniform probability between $0$ and $2\pi$\footnote{The PDF leading to \re{apic} is ${\cal P}_{AP}(|\delta^{(0)}_\bk|,\theta_\bk)=(2 \pi)^{-1} \delta_D\left(|\delta^{(0)}_\bk|-\sqrt{V P^0(k)}  \right)$, while for the gaussian distribution it is ${\cal P}_{Gauss}(|\delta^{(0)}_\bk|,\theta_\bk)=|\delta^{(0)}_\bk|/(\pi VP^0(k) ) \exp\left(-|\delta^{(0)}_\bk|^2/VP^0(k) \right)$ }.
The ensemble expectation values of the linear fields in the $AP$ distribution are zero for any product of odd fields, while for the even ones we have
\beqra
&&\!\!\!\!\!\!\!\! \!\!\!\!\!\!\!\! \!\!\!\!\!\!\!\! \!\!\!\!\!\!\!\!  \langle \delta^{(0)}_\bk \delta^{(0)}_{\bk'} \rangle_{AP} = V \sqrt{ P^0(k)\,P^0(k')} \;\langle e^{i(\theta_\bk+\theta_{\bk'})} \rangle_{AP} =  (2\pi)^3\,P^0(k) \delta_{\bk,-\bk'}\,,\nonumber\\
&&\!\!\!\!\!\!\!\! \!\!\!\!\!\!\!\! \!\!\!\!\!\!\!\! \!\!\!\!\!\!\!\!  \langle \delta^{(0)}_\bk \delta^{(0)}_{\bk'} \delta^{(0)}_{\bk''}\delta^{(0)}_{\bk'''} \rangle_{AP} = V^{2} \sqrt{ P^0(k)\,P^0(k')\,P^0(k'')\,P^0(k''')} \;\langle e^{i(\theta_\bk+\theta_{\bk'}+\theta_{\bk''}+\theta_{\bk'''})} \rangle_{AP}\,,\nonumber\\
&&\!\!\!\!\!\!\!\! \!\!\!\!\!\!\!\! \!\!\!\!\!\!\!\! \!\!\!\!\!\!\!\! =(2\pi)^6 \bigg( \delta_{\bk,-\bk'}\delta_{\bk'',-\bk'''} P^0(k) P^0(k'') +  \delta_{\bk,-\bk''}\delta_{\bk',-\bk'''} P^0(k) P^0(k') \,\nonumber\\
&&+ \delta_{\bk,-\bk'''}\delta_{\bk',-\bk''} P^0(k) P^0(k') \bigg) \nonumber\\
&&\!\!\!\!\!\!\!\! \!\!\!\!\!\!\!\! \!\!\!\!\!\!\!\!  -  \frac{(2\pi)^9 }{V} (P^0(k))^2\bigg( \delta_{\bk,\bk'}\delta_{\bk,-\bk''} \delta_{\bk,-\bk'''} +
  \delta_{\bk,\bk''}\delta_{\bk,-\bk'} \delta_{\bk,-\bk'''} + \delta_{\bk,\bk'''}\delta_{\bk,-\bk'} \delta_{\bk,-\bk''}\bigg)\,, \nonumber\\
&& \cdots
\label{D5}
\eeqra
While the two-point correlator is the same as for the gaussian distribution, for the four-point one the AP distribution gives the extra contribution at the last line, which represents the violaton of Wick's theorem due to the non-gaussianity of the AP PDF (notice that our results differs by a factor 2 with respect to eq. (10) in \cite{Angulo:2016hjd}). \\
This contribution is exactly what sets to zero the $O( (P^0(k))^2)$ contributions to the covariance of the PS,
\beqra
&&\!\!\!\!\!\!\!\! \!\!\!\!\!\!\!\! \!\!\!\!\!\!\!\!\cov^P_{AP,mn}=\langle \hat P_m\hat P_n\rangle_{AP} - \langle \hat P_m\rangle_{AP}\langle\hat P_n\rangle_{AP}\nonumber\\
&&\!\!\!\!\!\!\!\! \!\!\!\!\!\!\!\! \!\!\!\!\!\!\!\!\frac{1}{V} \int_{k_m} \frac{d^3 k}{V_s(k_m)}  \frac{1}{V} \int_{k_n} \frac{d^3 k'}{V_s(k_n)}   \bigg( \langle \delta_\bk\delta_{-\bk}  \delta_{\bk'}\delta_{-\bk'} \rangle_{AP}- \langle \delta_\bk\delta_{-\bk} \rangle_{AP}\langle \delta_{\bk'}\delta_{-\bk'} \rangle_{AP}\bigg)\nonumber\\
&&\!\!\!\!\!\!\!\! \!\!\!\!\!\!\!\! \!\!\!\!\!\!\!\!\frac{1}{V} \int_{k_m} \frac{d^3 k}{V_s(k_m)}  \frac{1}{V} \int_{k_n} \frac{d^3 k'}{V_s(k_n)}  V^2 \bigg( P^0(k)P^0(k') -P^0(k)P^0(k') \bigg) + O( (P^0(k))^3)\,,\nonumber\\
&&\!\!\!\!\!\!\!\! \!\!\!\!\!\!\!\! \!\!\!\!\!\!\!\!=O( (P^0(k))^3)\,.
\label{covdef2}
\eeqra\\
 Using the compact formalism for the vertices and fields (that is $\delta_k=e^\eta \vp_{1,\bk}$, $-\theta/{\cal H} f= e^\eta \vp_{2,\bk}$, with $\eta = \log D_+$), we rewrite \re{covdef2} as
\beqra
\cov^P_{AP,mn}&=&\frac{e^{4\eta}}{V^2} \int_{k_m} \frac{d^3 k}{V_s(k_m)} \int_{k_n} \frac{d^3 k}{V_s(k_n)} \bigg( \langle \vp_{1,\bk} \vp_{1,-\bk}\vp_{1,\bk'}\vp_{1,-\bk'} \rangle_{AP} \nonumber\\
&&-\langle \vp_{1,\bk} \vp_{1,-\bk}\rangle_{AP}\langle\vp_{1,\bk'}\vp_{1,-\bk'} \rangle_{AP}\bigg)\,.
\eeqra
As we have seen, the quantity inside parentheses vanishes at lowest PT order ($O((P^0)^2))$,
At $O((P^0)^3)$, the first term in parentheses has the following contributions
\beqra
&&\!\!\!\!\!\!\!\!  \langle \vp_{1,\bk} \vp_{1,-\bk}\vp_{1,\bk'}\vp_{1,-\bk'} \rangle_{AP} \to \langle1100 \rangle_{AP}+ \langle1010 \rangle_{AP}+ \langle1001 \rangle_{AP}+ \langle0110 \rangle_{AP}\nonumber\\
&&\!\!\!\!\!\!\!\!+ \langle0101 \rangle_{AP}+ \langle0011 \rangle_{AP}+ \langle2000 \rangle_{AP}+ \langle0200 \rangle_{AP}+ \langle0020 \rangle_{AP}+ \langle0002 \rangle_{AP}\,,
\eeqra
where the $``1"$  and the  $``2"$ stand for 
\beqra
&&``1"\to \vp^{(1)}_{a,\bk} =e^\eta g_{ab}^{(1)} I_{\bk,\bq_1,\bq_2} \gamma_{bcd}(\bq_1,\bq_2) u_c u_d  \vp^{(0)}_{\bq_1} \vp^{(0)}_{\bq_2}\,,\nonumber\\
&&``2"\to \vp^{(2)}_{a,\bk} =2 e^{2\eta}g_{ab}^{(2)} I_{\bk,\bq_1,\bq_2} \gamma_{bcd}(\bq_1,\bq_2) u_d  \vp^{(1)}_{c,\bq_1} \vp^{(0)}_{\bq_2}\,,
\eeqra
respectively, where we have set the linear field in the linear growing mode, $\vp^{(0)}_{a,\bk}= \vp^{(0)}_{\bk} u_a$ (with $u_1=u_2=1$), and
\beq
g_{ab}^{(n)}= \frac{1}{n}\left(\begin{array}{cc}
3/5 &2/5\\
3/5&2/5\\
\end{array}\right)+  \frac{2}{5+2 n}\left(\begin{array}{cc}
2/5 &-2/5\\
-3/5&3/5\\
\end{array}
\right)\,.
\eeq
The subtracted part contributes with
\beqra
&&\!\!\!\!\!\!\!\! \langle \vp_{1,\bk} \vp_{1,-\bk}\rangle_{AP} \langle \vp_{1,\bk'}\vp_{1,-\bk'} \rangle_{AP} \to \langle 11\rangle_{AP}\langle 00 \rangle_{AP}+ \langle 00\rangle_{AP}\langle 11 \rangle_{AP}\nonumber\\
&&\!\!\!\!\!\!\!\! + 2 \langle 20\rangle_{AP}\langle 00 \rangle_{AP}+2 \langle 00\rangle_{AP}\langle 20 \rangle_{AP}\,.
\eeqra
Now, taking into account that for $``AP''$ expectation values (differently from gaussian ones) the following relations hold
\beqra
&& \langle1100 \rangle_{AP} = \langle 11\rangle_{AP}\langle 00 \rangle_{AP}\,,\qquad  \langle0011 \rangle_{AP} = \langle 00\rangle_{AP}\langle 11 \rangle_{AP}\,\nonumber\\
&&  \langle2000 \rangle_{AP} =  \langle0200 \rangle_{AP} =  \langle 20\rangle_{AP}\langle 00  \rangle_{AP}\,,
\eeqra
whe have that the surviving contributions are
\beqra
&& \bigg( \langle \vp_{1,\bk} \vp_{1,-\bk}\vp_{1,\bk'}\vp_{1,-\bk'} \rangle_{AP}  -\langle \vp_{1,\bk} \vp_{1,-\bk}\rangle_{AP}\langle\vp_{1,\bk'}\vp_{1,-\bk'} \rangle_{AP}\bigg) =\nonumber\\
&&\langle1010 \rangle_{AP}+ \langle1001 \rangle_{AP}+ \langle0110 \rangle_{AP} + \langle0101 \rangle_{AP} + O((P^0)^4)\,\nonumber\\
&& = \langle \vp^{(1)}_{1,\bk} \vp^{(0)}_{1,-\bk}\vp^{(1)}_{1,\bk'}\vp^{(0)}_{1,-\bk'} \rangle_{AP}+  \langle \vp^{(1)}_{1,\bk} \vp^{(0)}_{1,-\bk}\vp^{(0)}_{1,\bk'}\vp^{(1)}_{1,-\bk'} \rangle_{AP}\nonumber\\
&& +  \langle \vp^{(0)}_{1,\bk} \vp^{(1)}_{1,-\bk}\vp^{(1)}_{1,\bk'}\vp^{(0)}_{1,-\bk'} \rangle_{AP}+ \langle \vp^{(0)}_{1,\bk} \vp^{(1)}_{1,-\bk}\vp^{(0)}_{1,\bk'}\vp^{(1)}_{1,-\bk'} \rangle_{AP}+ O((P^0)^4)\,.
\eeqra
The non-vanishing contributions to the first term are given explicitly by
\beqra
&&\!\!\!\!\!\!\!\!\!\!\! \!\!\!\! \!\!\! \!\!\! \!\!\!\langle \vp^{(1)}_{1,\bk} \vp^{(0)}_{1,-\bk}\vp^{(1)}_{1,\bk'}\vp^{(0)}_{1,-\bk'} \rangle_{AP}= e^{2\eta}g_{1a}^{(1)}g_{1b}^{(1)} I_{\bk,\bq_1,\bq_2}I_{\bk',\bp_1,\bp_2}\tilde\gamma_a(\bq_1,\bq_2)\tilde\gamma_b(\bp_1,\bp_2)  \nonumber\\
&&\qquad\qquad\quad\times \langle\vp^{(0)}_{\bq_1}\vp^{(0)}_{\bq_2}\vp^{(0)}_{-\bk}\vp^{(0)}_{\bp_1}\vp^{(0)}_{\bp_2}\vp^{(0)}_{-\bk'} \rangle_{AP}\,,\nonumber\\
&&\!\!\!\!\!\!\!\!\!\!\! \!\!\!\! \!\!\! \!\!\! \!\!\! =e^{2\eta}(2\pi)^3 g_{1a}^{(1)}g_{1b}^{(1)} \int d^3 q_1 d^3 q_2 d^3 p_1 d^3 p_2\;\tilde\gamma_a(\bq_1,\bq_2)\tilde\gamma_b(\bp_1,\bp_2) P^0(k)P^0(q_1)P^0(q_2)\nonumber\\
&&\times\delta_D(\bk-\bq_1-\bq_2)\,\delta_D(\bk'-\bp_1-\bp_2)\nonumber\\
&&\!\!\!\!\!\!\!\!\!\!\! \!\!\!\! \!\!\! \!\!\! \!\!\! \times \Bigg\{ \delta_D(\bk+\bk' )\Big[\delta_D(\bq_1+\bp_1 )\delta_D(\bq_2+\bp_2 )+\delta_D(\bq_1+\bp_2 )\delta_D(\bq_2+\bp_1 ) \nonumber\\
&&\!\!\!\!\!\!\!\!\!\!\! \!\!\!\! \!\!\! \!\!\! \!\!\!  -  \frac{(2\pi)^3}{V}\delta_D(\bq_1+\bp_1 )\delta_D(\bq_2+\bp_2 )    \delta_D(\bq_1-\bq_2 )  \Big]\nonumber\\
&&\!\!\!\!\!\!\!\!\!\!\! \!\!\!\! \!\!\! \!\!\! \!\!\! +\delta_D(\bq_1-\bk')\Big[\delta_D(\bq_2+\bp_1 )\delta_D(-\bk+\bp_2 )+\delta_D(\bq_2+\bp_2 )\delta_D(-\bk+\bp_1 ) \Big]\nonumber\\
&&\!\!\!\!\!\!\!\!\!\!\! \!\!\!\! \!\!\! \!\!\! \!\!\! +\delta_D(\bq_2-\bk')\Big[\delta_D(\bq_1+\bp_1 )\delta_D(-\bk+\bp_2 )+\delta_D(\bq_1+\bp_2 )\delta_D(-\bk+\bp_1 ) \Big]\Bigg\}\,,
\eeqra
where the subtracted term at the second line of the last equation comes from the difference between the $``AP"$ averaging and the gaussian one, see eq.~\re{D5}, and we have defined $\tilde\gamma_a(\bq_1,\bq_2)\equiv \gamma_{abc}(\bq_1,\bq_2)u_a u_b$. Working out the delta functions and using $(2 \pi)^3 \delta_D(0) = V$ we get
\beqra
&&\!\!\!\!\!\!\!\!\!\!\! \!\!\!\! \!\!\! \!\!\! \!\!\! \langle \vp^{(1)}_\bk \vp^{(0)}_{-\bk}\vp^{(1)}_{\bk'}\vp^{(0)}_{-\bk'} \rangle_{AP}=  (2 \pi)^3 e^{-4\eta} \delta_D(\bk+\bk') \left[V P^0(k) \Delta P_{22}(k) -4P^0(k) (P^0(k/2))^2 \right]\nonumber\\
&&\qquad\qquad+4e^{-4\eta} V F_2(\bk',\bk-\bk') F_2(\bk,\bk'-\bk) P^0(k)P^0(k') P^0(|\bk-\bk'|)\,,\nonumber\\
\eeqra
where we have used $F_2(\bk',\bk-\bk') = g_{1a}^{(1)} \tilde\gamma_{a}(\bk',\bk-\bk')$, and $ \Delta P_{22}(k)$ is the $``22"$ contribution to the 1-loop PS \cite{PT}.\\
Summing up the four contributions we finally have
\beqra
&&\!\!\!\!\!\!\!\!\!\!\! \!\!\!\! \!\!\! \!\!\! \!\!\!\cov^P_{AP,mn}= \frac{(2\pi)^3}{V}\Bigg[ \left(4\frac{ P^0(k) \Delta P_{22}(k)}{V_s(k_m)} -16 \frac{P^0(k)(P^0(k/2))^2}{V\,V_s(k_m)} \right)\delta_{mn}+ \frac{1}{(2\pi)^3} T^{AP}_{mn}\Bigg]\,,\nonumber\\
\label{covan}
\eeqra
where
\beqra
&&\!\!\!\!\!\!\!\!\!\!\! \!\!\!\! \!\!\! \!\!\! \!\!\!T^{AP}_{mn}\equiv 8 \int_{k_m} \frac{d^3 k}{V_s(k_m)}\int_{k_n} \frac{d^3 k'}{V_s(k_n)} \Bigg( F_2(\bk',\bk-\bk') F_2(\bk,\bk'-\bk)P^0(k)P^0(k') P^0(|\bk-\bk'|)\nonumber\\
&&\qquad\qquad\qquad\qquad\qquad+ (\bk'\to - \bk')\Bigg)\,.
\eeqra

As we see, the diagonal contribution to the covariance matrix is only $O((P^0)^3)$ and is therefore suppressed with respect to that obtained in the standard method with gaussian initial conditions.

\begin{figure}
\includegraphics[width=0.8\textwidth,right]{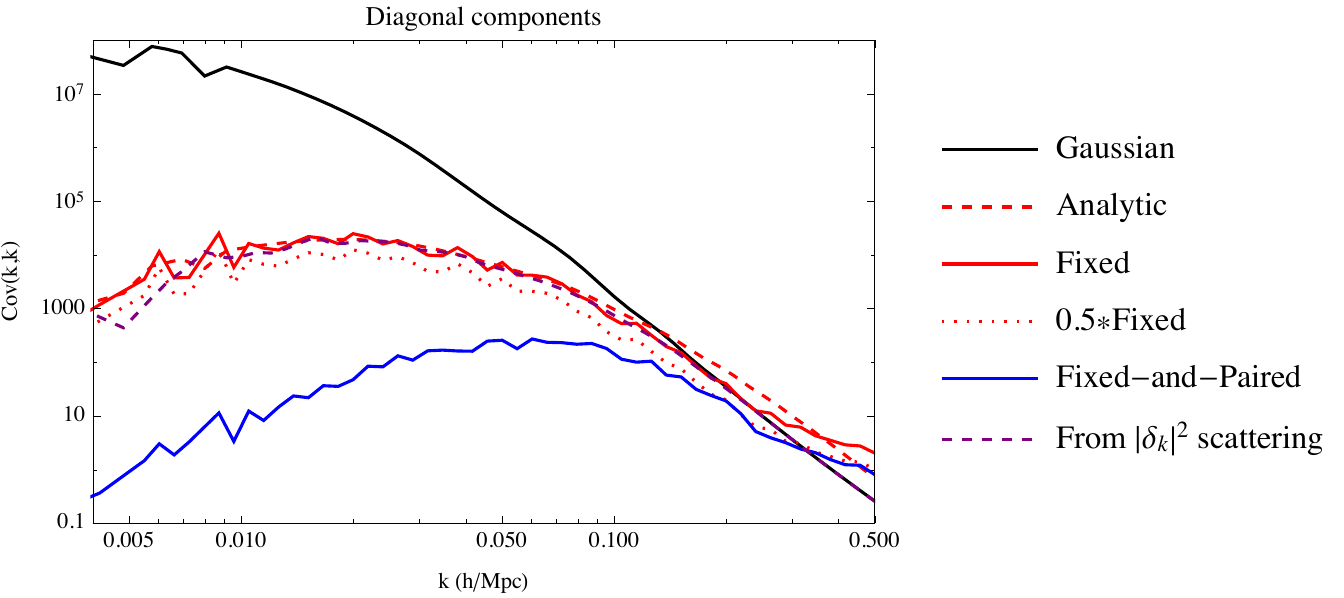}
\caption{The diagonal entries of the PS covariance matrix as obtained in the standard method with gaussian initial conditions (black solid) and in the AP method, as measured from N-body simulations in the ``Fixed" approach (red-solid) and in the ``Fixed-and-Paired" one (blue solid). We also show the estimate from the scattering of $|\delta_{\vec{k}}|^2$ for $\vec{k}$'s in each of the k bins (purple-dashed) and from  the analytic result of eq.~\re{covan} (red-dashed).  }
\label{comparecov}
\end{figure}

 This is shown explicitely in Fig.~\ref{comparecov}, where we see that our analytic computation reproduces, at small $k's$, the result of the measurement of the covariance matrix from the scattering of $|\delta_{\vec{k}}|^2$ for $\vec{k}$'s in each of the k bins.
 
 We also plot the results obtained via N-body simulations by performing $100$ AP pairs (i.e., $200$ simulations) with the same simulation parameters except a much smaller number of particles $256^3$. We analyse the simulated data only at $z=0$ for the matter field. These simulations allow us to estimate the covariance matrix including the non-diagonal entries both for the paired and unpaired cases. While the resolution is much poorer and the number of realizations might be small for the estimation of the covariance matrix, this new set of simulations would be helpful to check the expectations from the analytical argument at least qualitatively.  In Fig.~\ref{comparecov} we confirm the suppression of the variance for the AP simulations on large scales compared with that from Gaussian initial conditions. The fixed-and-paired method is especially efficient to reduce the variance on scales $k\lesssim 0.1\,h\mathrm{Mpc}^{-1}$. On smaller scales, this approaches to the half of the fixed case. This shows that the cancellation is no longer effective, and the variance is reduced simply because we have doubled the number of Fourier mode by using two simulations. The covariance eventually exceeds the Gaussian value on smaller scales, where we see that the paring no longer helps. We also see that the analytic result compares rather well with the numerical one for the fixed method for $k\alt 0.1\;\mathrm{h\,Mpc^{-1}}$.

\begin{figure}
\includegraphics[width=0.8\textwidth,right]{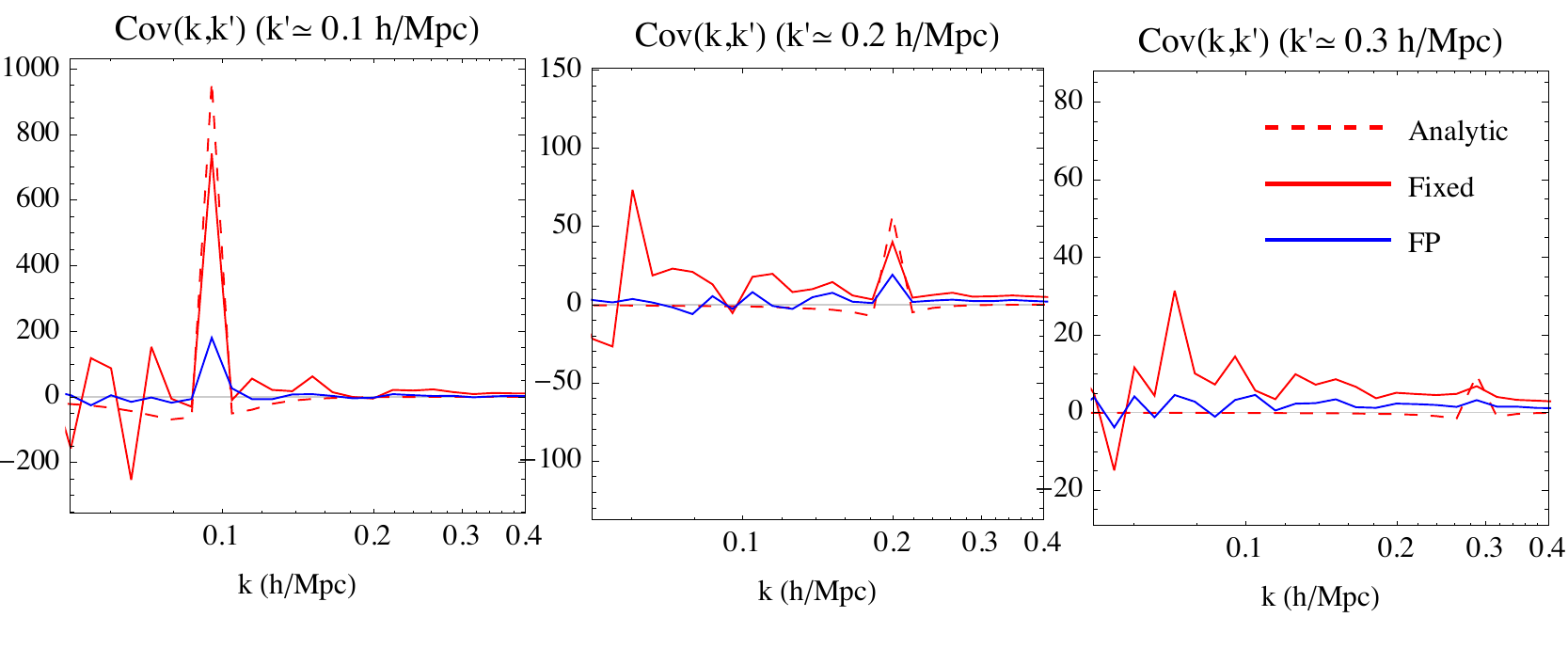}
\includegraphics[width=0.8\textwidth,right]{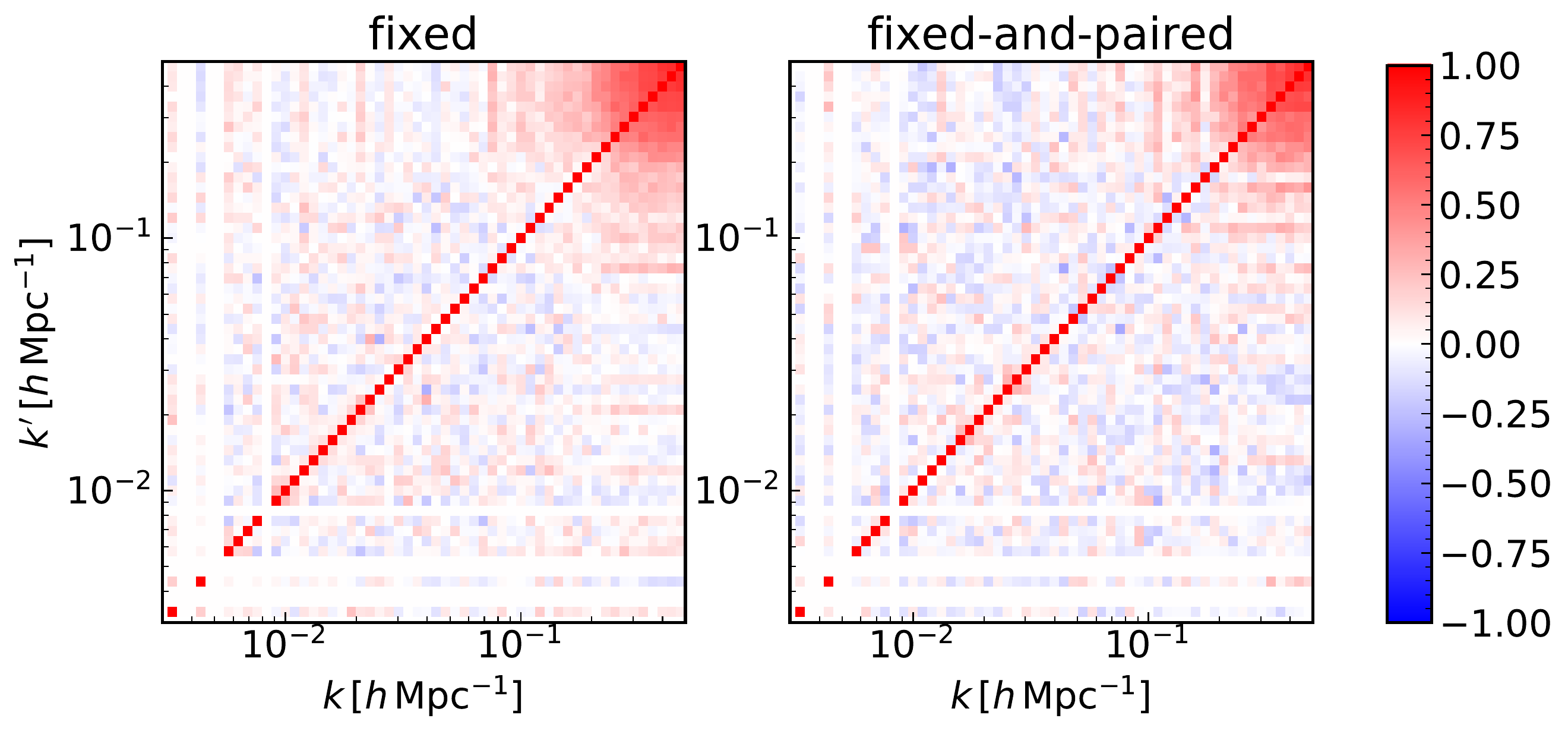}
\includegraphics[width=0.8\textwidth,right]{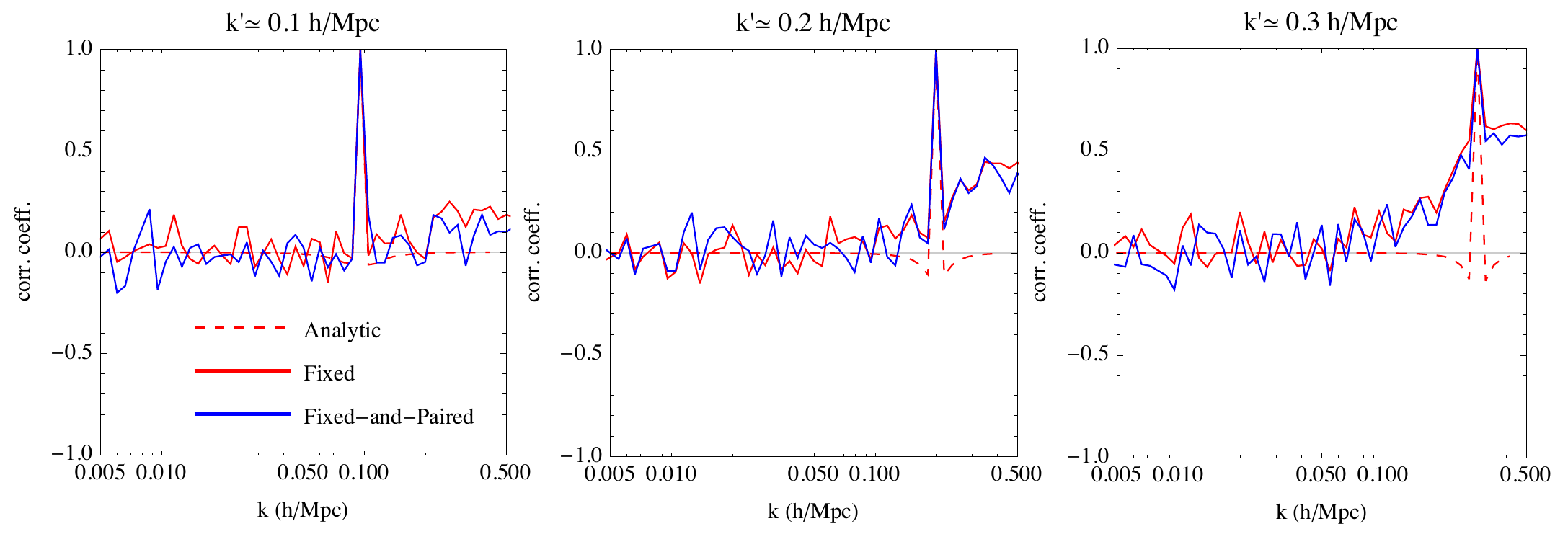}
\caption{Estimated covariance matrix $\cov^P_{AP}(k,k')$ for some fixed values of $k'$  (upper panel), the matrix of correlation coefficients, $R(k,k')=\cov^P_{AP}(k,k')/\sqrt{\cov^P_{AP}(k,k)\cov^P_{AP}(k',k')}$ (central panel) and the same quantity as a function of $k$ for some fixed values of $k'$ (lower panel).}
\label{cov_sim}
\end{figure}

In Fig.~\ref{cov_sim} we  show the full covariance matrix as measured in simulations, the non-diagonal entries $\cov(k,k')$ for some fixed value of one of the two momenta, and the  cross-correlation coefficient, $\mathrm{cov}(k,k')/\sqrt{\mathrm{cov}(k,k)\,\mathrm{cov}(k',k')}$, to see the relative importance of the off-diagonal entries. The general trend looks very similar in the two cases, and we confirm that the numerical and the analytical results agree at large scales.

\begin{figure}
\includegraphics[width=0.8\textwidth,right]{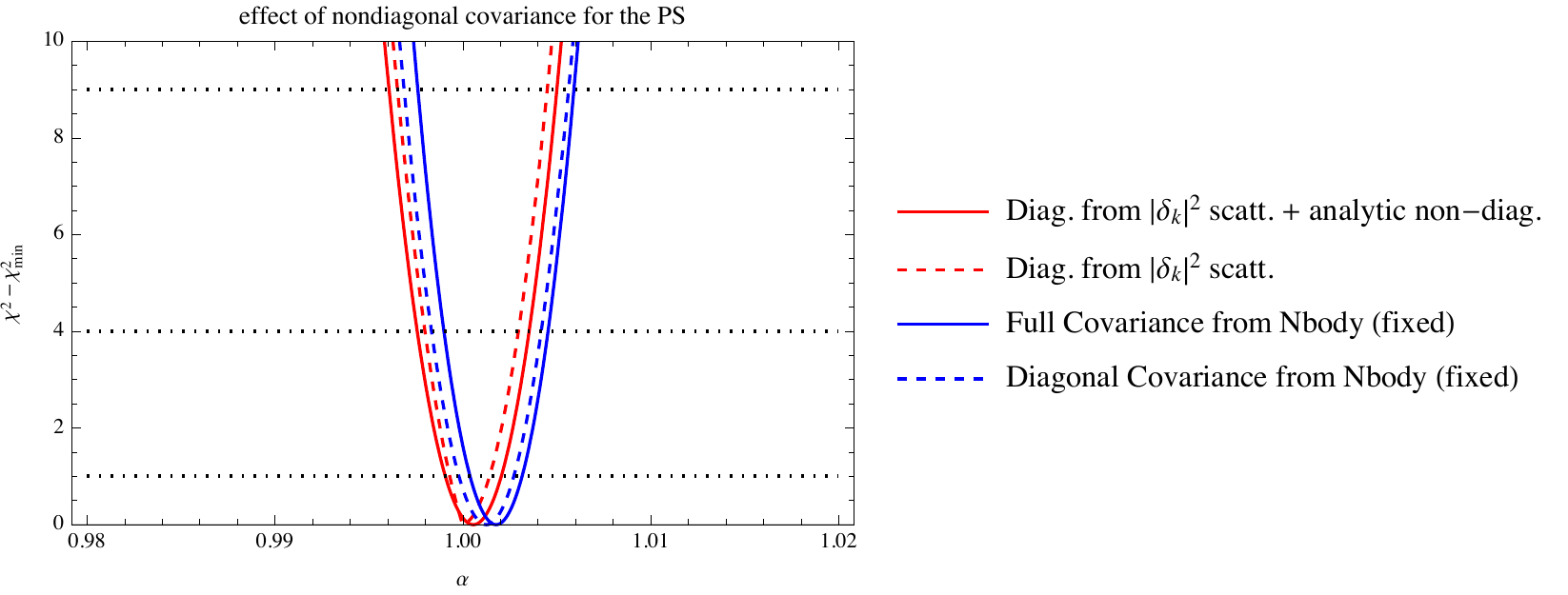}
\caption{The effect of the non-diagonal entries of the covariance matrix for the PS on the fitting procedures discussed in the paper. We compare the extractors from N-Body simulations and from the TRG for DM at $z=0$. To compute the red lines we estimated, as in the rest of the paper, the diagonal entries of the PS covariance matrix from the scattering of $|\delta_{\vec k}|^2$ inside each momentum bin. In the continuous red line we added the effect of non-diagonal entries computed analytically using eq.~\re{covan}. The blue lines are derived using the PS covariance measured from simulations. }
\label{nondiagcov}
\end{figure}

 Finally, we added  the non-diagonal entries of the PS covariance matrix in the covariance matrix for the extractor, eq.~\re{DRdisc}, to see their impact on our $\chi^2$ analyses. This is shown in Fig.~\ref{nondiagcov}, where we show, by comparison, the effect of including these terms when fitting the TRG PS to the one extracted from the simulations for DM at $z=0$. As anticipated, the effect of the non-diagonal terms is to all extents negligible, and therefore we only considered the diagonal covariance matrix for the PS in the paper.  Notice that, coherently with what we have done in the rest of the paper, we have used the non-diagonal entries of the PS covariance matrix for the ``fixed'' method (both in the analytical and in the numerical determinations), which as seen in Figs. \ref{comparecov}, \ref{cov_sim}, gives a conservative estimate of the covariance of our AP simulations obtained with the fixed-and-paired method.  



\section*{References}
\bibliographystyle{JHEP}
\bibliography{/Users/massimo/Bibliografia/mybib.bib}

\providecommand{\href}[2]{#2}\begingroup\raggedright\begin{thebibliography}{10}

\bibitem{Eis05}
{\bf SDSS} Collaboration, D.~J. Eisenstein {\em et.~al.}, {\it {Detection of
  the Baryon Acoustic Peak in the Large-Scale Correlation Function of SDSS
  Luminous Red Galaxies}},  {\em Astrophys. J.} {\bf 633} (2005) 560--574
  [\href{http://arXiv.org/abs/astro-ph/0501171}{{\tt astro-ph/0501171}}].

\bibitem{Cole:2005sx}
{\bf 2dFGRS} Collaboration, S.~Cole {\em et.~al.}, {\it {The 2dF Galaxy
  Redshift Survey: Power-spectrum analysis of the final dataset and
  cosmological implications}},  {\em Mon.Not.Roy.Astron.Soc.} {\bf 362} (2005)
  505--534 [\href{http://arXiv.org/abs/astro-ph/0501174}{{\tt
  astro-ph/0501174}}].

\bibitem{Eisenstein:2006nj}
D.~J. Eisenstein, H.-j. Seo and .~White, Martin~J., {\it {On the Robustness of
  the Acoustic Scale in the Low-Redshift Clustering of Matter}},  {\em
  Astrophys.J.} {\bf 664} (2007) 660--674
  [\href{http://arXiv.org/abs/astro-ph/0604361}{{\tt astro-ph/0604361}}].

\bibitem{PT}
F.~Bernardeau, S.~Colombi, E.~Gaztanaga and R.~Scoccimarro, {\it {Large-scale
  structure of the universe and cosmological perturbation theory}},  {\em Phys.
  Rept.} {\bf 367} (2002) 1--248
  [\href{http://arXiv.org/abs/astro-ph/0112551}{{\tt astro-ph/0112551}}].

\bibitem{RPTb}
M.~Crocce and R.~Scoccimarro, {\it {Memory of Initial Conditions in
  Gravitational Clustering}},  {\em Phys. Rev.} {\bf D73} (2006) 063520
  [\href{http://arXiv.org/abs/astro-ph/0509419}{{\tt astro-ph/0509419}}].

\bibitem{MP07b}
S.~Matarrese and M.~Pietroni, {\it {Resumming Cosmic Perturbations}},  {\em
  JCAP} {\bf 0706} (2007) 026
  [\href{http://arXiv.org/abs/astro-ph/0703563}{{\tt astro-ph/0703563}}].

\bibitem{RPTBAO}
M.~Crocce and R.~Scoccimarro, {\it {Nonlinear Evolution of Baryon Acoustic
  Oscillations}},  {\em Phys. Rev.} {\bf D77} (2008) 023533
  [\href{http://arXiv.org/abs/0704.2783}{{\tt 0704.2783}}].

\bibitem{Pietroni08}
M.~Pietroni, {\it {Flowing with Time: a New Approach to Nonlinear Cosmological
  Perturbations}},  {\em JCAP} {\bf 0810} (2008) 036
  [\href{http://arXiv.org/abs/0806.0971}{{\tt 0806.0971}}].

\bibitem{Bernardeau:2008fa}
F.~Bernardeau, M.~Crocce and R.~Scoccimarro, {\it {Multi-Point Propagators in
  Cosmological Gravitational Instability}},  {\em Phys.Rev.} {\bf D78} (2008)
  103521 [\href{http://arXiv.org/abs/0806.2334}{{\tt 0806.2334}}].

\bibitem{Taruya:2009ir}
A.~Taruya, T.~Nishimichi, S.~Saito and T.~Hiramatsu, {\it {Non-linear Evolution
  of Baryon Acoustic Oscillations from Improved Perturbation Theory in Real and
  Redshift Spaces}},  {\em Phys.Rev.} {\bf D80} (2009) 123503
  [\href{http://arXiv.org/abs/0906.0507}{{\tt 0906.0507}}].

\bibitem{Senatore:2014via}
L.~Senatore and M.~Zaldarriaga, {\it {The IR-resummed Effective Field Theory of
  Large Scale Structures}},  {\em JCAP} {\bf 1502} (2015), no.~02 013
  [\href{http://arXiv.org/abs/1404.5954}{{\tt 1404.5954}}].

\bibitem{Blas:2016sfa}
D.~Blas, M.~Garny, M.~M. Ivanov and S.~Sibiryakov, {\it {Time-Sliced
  Perturbation Theory II: Baryon Acoustic Oscillations and Infrared
  Resummation}},  {\em JCAP} {\bf 1607} (2016), no.~07 028
  [\href{http://arXiv.org/abs/1605.02149}{{\tt 1605.02149}}].

\bibitem{Peloso:2016qdr}
M.~Peloso and M.~Pietroni, {\it {Galilean invariant resummation schemes of
  cosmological perturbations}},  {\em JCAP} {\bf 1701} (2017), no.~01 056
  [\href{http://arXiv.org/abs/1609.06624}{{\tt 1609.06624}}].

\bibitem{Baumann:2010tm}
D.~Baumann, A.~Nicolis, L.~Senatore and M.~Zaldarriaga, {\it {Cosmological
  Non-Linearities as an Effective Fluid}},  {\em JCAP} {\bf 1207} (2012) 051
  [\href{http://arXiv.org/abs/1004.2488}{{\tt 1004.2488}}].

\bibitem{Pietroni:2011iz}
M.~Pietroni, G.~Mangano, N.~Saviano and M.~Viel, {\it {Coarse-Grained
  Cosmological Perturbation Theory}},  {\em JCAP} {\bf 1201} (2012) 019
  [\href{http://arXiv.org/abs/1108.5203}{{\tt 1108.5203}}].

\bibitem{Carrasco:2012cv}
J.~J.~M. Carrasco, M.~P. Hertzberg and L.~Senatore, {\it {The Effective Field
  Theory of Cosmological Large Scale Structures}},  {\em JHEP} {\bf 1209}
  (2012) 082 [\href{http://arXiv.org/abs/1206.2926}{{\tt 1206.2926}}].

\bibitem{Manzotti:2014loa}
A.~Manzotti, M.~Peloso, M.~Pietroni, M.~Viel and F.~Villaescusa-Navarro, {\it
  {A coarse grained perturbation theory for the Large Scale Structure, with
  cosmology and time independence in the UV}},  {\em JCAP} {\bf 1409} (2014),
  no.~09 047 [\href{http://arXiv.org/abs/1407.1342}{{\tt 1407.1342}}].

\bibitem{Blas:2015tla}
D.~Blas, S.~Floerchinger, M.~Garny, N.~Tetradis and U.~A. Wiedemann, {\it
  {Large scale structure from viscous dark matter}},  {\em JCAP} {\bf 1511}
  (2015) 049 [\href{http://arXiv.org/abs/1507.06665}{{\tt 1507.06665}}].

\bibitem{Floerchinger:2016hja}
S.~Floerchinger, M.~Garny, N.~Tetradis and U.~A. Wiedemann, {\it
  {Renormalization-group flow of the effective action of cosmological
  large-scale structures}},  {\em JCAP} {\bf 1701} (2017), no.~01 048
  [\href{http://arXiv.org/abs/1607.03453}{{\tt 1607.03453}}].

\bibitem{Noda:2017tfh}
E.~Noda, M.~Peloso and M.~Pietroni, {\it {A Robust BAO Extractor}},
  \href{http://arXiv.org/abs/1705.01475}{{\tt 1705.01475}}.

\bibitem{Eisenstein:2006nk}
D.~J. Eisenstein, H.-j. Seo, E.~Sirko and D.~Spergel, {\it {Improving
  Cosmological Distance Measurements by Reconstruction of the Baryon Acoustic
  Peak}},  {\em Astrophys.J.} {\bf 664} (2007) 675--679
  [\href{http://arXiv.org/abs/astro-ph/0604362}{{\tt astro-ph/0604362}}].

\bibitem{Seo:2008yx}
H.-J. Seo, E.~R. Siegel, D.~J. Eisenstein and M.~White, {\it {Non-linear
  structure formation and the acoustic scale}},  {\em Astrophys.J.} {\bf 686}
  (2008) 13--24 [\href{http://arXiv.org/abs/0805.0117}{{\tt 0805.0117}}].

\bibitem{Padmanabhan:2008dd}
N.~Padmanabhan, M.~White and J.~Cohn, {\it {Reconstructing Baryon Oscillations:
  A Lagrangian Theory Perspective}},  {\em Phys.Rev.} {\bf D79} (2009) 063523
  [\href{http://arXiv.org/abs/0812.2905}{{\tt 0812.2905}}].

\bibitem{Noh:2009bb}
Y.~Noh, M.~White and N.~Padmanabhan, {\it {Reconstructing baryon
  oscillations}},  {\em Phys.Rev.} {\bf D80} (2009) 123501
  [\href{http://arXiv.org/abs/0909.1802}{{\tt 0909.1802}}].

\bibitem{Tassev:2012hu}
S.~Tassev and M.~Zaldarriaga, {\it {Towards an Optimal Reconstruction of Baryon
  Oscillations}},  {\em JCAP} {\bf 1210} (2012) 006
  [\href{http://arXiv.org/abs/1203.6066}{{\tt 1203.6066}}].

\bibitem{White:2015eaa}
M.~White, {\it {Reconstruction within the Zeldovich approximation}},  {\em Mon.
  Not. Roy. Astron. Soc.} {\bf 450} (2015), no.~4 3822--3828
  [\href{http://arXiv.org/abs/1504.03677}{{\tt 1504.03677}}].

\bibitem{Anderson:2013zyy}
{\bf BOSS} Collaboration, L.~Anderson {\em et.~al.}, {\it {The clustering of
  galaxies in the SDSS-III Baryon Oscillation Spectroscopic Survey: baryon
  acoustic oscillations in the Data Releases 10 and 11 Galaxy samples}},  {\em
  Mon.Not.Roy.Astron.Soc.} {\bf 441} (2014), no.~1 24--62
  [\href{http://arXiv.org/abs/1312.4877}{{\tt 1312.4877}}].

\bibitem{Peloso:2015jua}
M.~Peloso, M.~Pietroni, M.~Viel and F.~Villaescusa-Navarro, {\it {The effect of
  massive neutrinos on the BAO peak}},  {\em JCAP} {\bf 1507} (2015), no.~07
  001 [\href{http://arXiv.org/abs/1505.07477}{{\tt 1505.07477}}].

\bibitem{Anselmi:2017cuq}
S.~Anselmi, G.~D. Starkman, P.-S. Corasaniti, R.~K. Sheth and I.~Zehavi, {\it
  {The Linear Point: A cleaner cosmological standard ruler}},
  \href{http://arXiv.org/abs/1703.01275}{{\tt 1703.01275}}.

\bibitem{Komatsu:2008hk}
{\bf WMAP} Collaboration, E.~Komatsu {\em et.~al.}, {\it {Five-Year Wilkinson
  Microwave Anisotropy Probe (WMAP) Observations: Cosmological
  Interpretation}},  {\em Astrophys. J. Suppl.} {\bf 180} (2009) 330--376
  [\href{http://arXiv.org/abs/0803.0547}{{\tt 0803.0547}}].

\bibitem{Sherwin:2012nh}
B.~D. Sherwin and M.~Zaldarriaga, {\it {The Shift of the Baryon Acoustic
  Oscillation Scale: A Simple Physical Picture}},  {\em Phys.Rev.} {\bf D85}
  (2012) 103523 [\href{http://arXiv.org/abs/1202.3998}{{\tt 1202.3998}}].

\bibitem{Desjacques:2008jj}
V.~Desjacques, {\it {Baryon acoustic signature in the clustering of density
  maxima}},  {\em Phys.Rev.} {\bf D78} (2008) 103503
  [\href{http://arXiv.org/abs/0806.0007}{{\tt 0806.0007}}].

\bibitem{Desjacques:2009kt}
V.~Desjacques and R.~K. Sheth, {\it {Redshift space correlations and
  scale-dependent stochastic biasing of density peaks}},  {\em Phys.Rev.} {\bf
  D81} (2010) 023526 [\href{http://arXiv.org/abs/0909.4544}{{\tt 0909.4544}}].

\bibitem{Baldauf:2014fza}
T.~Baldauf, V.~Desjacques and U.~Seljak, {\it {Velocity bias in the
  distribution of dark matter halos}},
  \href{http://arXiv.org/abs/1405.5885}{{\tt 1405.5885}}.

\bibitem{Scoccimarro:2004tg}
R.~Scoccimarro, {\it {Redshift-space distortions, pairwise velocities and
  nonlinearities}},  {\em Phys.Rev.} {\bf D70} (2004) 083007
  [\href{http://arXiv.org/abs/astro-ph/0407214}{{\tt astro-ph/0407214}}].

\bibitem{Eisenstein:1997ik}
D.~J. Eisenstein and W.~Hu, {\it {Baryonic features in the matter transfer
  function}},  {\em Astrophys. J.} {\bf 496} (1998) 605
  [\href{http://arXiv.org/abs/astro-ph/9709112}{{\tt astro-ph/9709112}}].

\bibitem{Nishimichi:2007xt}
T.~Nishimichi, H.~Ohmuro, M.~Nakamichi, A.~Taruya, K.~Yahata, A.~Shirata,
  S.~Saito, H.~Nomura, K.~Yamamoto and Y.~Suto, {\it {Characteristic Scales of
  Baryon Acoustic Oscillations from Perturbation Theory: Non-linearity and
  Redshift-Space Distortion Effects}},  {\em Publ. Astron. Soc. Jap.} {\bf 59}
  (2007) 1049 [\href{http://arXiv.org/abs/0705.1589}{{\tt 0705.1589}}].

\bibitem{LMPR09}
J.~Lesgourgues, S.~Matarrese, M.~Pietroni and A.~Riotto, {\it {Non-linear Power
  Spectrum including Massive Neutrinos: the Time-RG Flow Approach}},  {\em
  JCAP} {\bf 0906} (2009) 017 [\href{http://arXiv.org/abs/0901.4550}{{\tt
  0901.4550}}].

\bibitem{Senatore:2017hyk}
L.~Senatore and M.~Zaldarriaga, {\it {The Effective Field Theory of Large-Scale
  Structure in the presence of Massive Neutrinos}},
  \href{http://arXiv.org/abs/1707.04698}{{\tt 1707.04698}}.

\bibitem{Upadhye:2017hdl}
A.~Upadhye, {\it {Neutrino mass and dark energy constraints from redshift-space
  distortions}},  \href{http://arXiv.org/abs/1707.09354}{{\tt 1707.09354}}.

\bibitem{Peloso:2013zw}
M.~Peloso and M.~Pietroni, {\it {Galilean invariance and the consistency
  relation for the nonlinear squeezed bispectrum of large scale structure}},
  {\em JCAP} {\bf 1305} (2013) 031 [\href{http://arXiv.org/abs/1302.0223}{{\tt
  1302.0223}}].

\bibitem{Springel:2005mi}
V.~Springel, {\it {The Cosmological simulation code GADGET-2}},  {\em
  Mon.Not.Roy.Astron.Soc.} {\bf 364} (2005) 1105--1134
  [\href{http://arXiv.org/abs/astro-ph/0505010}{{\tt astro-ph/0505010}}].

\bibitem{Ade:2015xua}
{\bf Planck} Collaboration, P.~A.~R. Ade {\em et.~al.}, {\it {Planck 2015
  results. XIII. Cosmological parameters}},  {\em Astron. Astrophys.} {\bf 594}
  (2016) A13 [\href{http://arXiv.org/abs/1502.01589}{{\tt 1502.01589}}].

\bibitem{Nishimichi:2008ry}
T.~Nishimichi {\em et.~al.}, {\it {Modeling Nonlinear Evolution of Baryon
  Acoustic Oscillations: Convergence Regime of N-body Simulations and Analytic
  Models}},  {\em Publ. Astron. Soc. Jap.} {\bf 61} (2009) 321
  [\href{http://arXiv.org/abs/0810.0813}{{\tt 0810.0813}}].

\bibitem{Valageas:2010yw}
P.~Valageas and T.~Nishimichi, {\it {Combining perturbation theories with halo
  models}},  {\em Astron. Astrophys.} {\bf 527} (2011) A87
  [\href{http://arXiv.org/abs/1009.0597}{{\tt 1009.0597}}].

\bibitem{Scoccimarro:1997gr}
R.~Scoccimarro, {\it {Transients from initial conditions: a perturbative
  analysis}},  {\em Mon. Not. Roy. Astron. Soc.} {\bf 299} (1998) 1097
  [\href{http://arXiv.org/abs/astro-ph/9711187}{{\tt astro-ph/9711187}}].

\bibitem{Crocce:2006ve}
M.~Crocce, S.~Pueblas and R.~Scoccimarro, {\it {Transients from Initial
  Conditions in Cosmological Simulations}},  {\em Mon.Not.Roy.Astron.Soc.} {\bf
  373} (2006) 369--381 [\href{http://arXiv.org/abs/astro-ph/0606505}{{\tt
  astro-ph/0606505}}].

\bibitem{Taruya:2012ut}
A.~Taruya, F.~Bernardeau, T.~Nishimichi and S.~Codis, {\it {RegPT: Direct and
  fast calculation of regularized cosmological power spectrum at two-loop
  order}},  {\em Phys.Rev.} {\bf D86} (2012) 103528
  [\href{http://arXiv.org/abs/1208.1191}{{\tt 1208.1191}}].

\bibitem{Angulo:2016hjd}
R.~E. Angulo and A.~Pontzen, {\it {Cosmological $N$-body simulations with
  suppressed variance}},  {\em Mon. Not. Roy. Astron. Soc.} {\bf 462} (2016),
  no.~1 L1--L5 [\href{http://arXiv.org/abs/1603.05253}{{\tt 1603.05253}}].

\bibitem{Behroozi:2011ju}
P.~S. Behroozi, R.~H. Wechsler and H.-Y. Wu, {\it {The Rockstar Phase-Space
  Temporal Halo Finder and the Velocity Offsets of Cluster Cores}},  {\em
  Astrophys. J.} {\bf 762} (2013) 109
  [\href{http://arXiv.org/abs/1110.4372}{{\tt 1110.4372}}].

\bibitem{Scoccimarro:1999kp}
R.~Scoccimarro, M.~Zaldarriaga and L.~Hui, {\it {Power spectrum correlations
  induced by nonlinear clustering}},  {\em Astrophys. J.} {\bf 527} (1999) 1
  [\href{http://arXiv.org/abs/astro-ph/9901099}{{\tt astro-ph/9901099}}].

\end{thebibliography}\endgroup
\end{document}